\documentclass[12pt]{article}

\pdfoutput=1

\usepackage{amssymb,amsfonts,amsthm}
\usepackage{amsmath}
\usepackage{bm}             
\usepackage[mathscr]{eucal}
\usepackage{cancel}
\usepackage{wasysym}

\usepackage{graphicx}
\usepackage{subfigure}
\def\be{\begin{equation}}
\def\ee{\end{equation}}
\def\bea{\begin{eqnarray}}
\def\eea{\end{eqnarray}}

\def\bdelta{\mbox{\boldmath $\delta$}}

\def\hsp5{\hspace{5mm}}

\theoremstyle{remark}

\newcommand{\sfrac}[2]{{\textstyle{#1\over#2}}}

\title{\sc Perturbations of the Lambda-CDM model in a dynamical
systems perspective}

\begin{document}

\author{
\sc Artur Alho,$^{1}$\thanks{Electronic address:{\tt
artur.alho@tecnico.ulisboa.pt}}\,, Claes Uggla,$^{2}$\thanks{Electronic address:{\tt
claes.uggla@kau.se}}\,  and John Wainwright$^{3}$\thanks{Electronic
address:{\tt jwainwri@uwaterloo.ca}}\\
$^{1}${\small\em Center for Mathematical Analysis, Geometry and Dynamical Systems,}\\
{\small\em Instituto Superior T\'ecnico, Universidade de Lisboa,}\\
{\small\em Av. Rovisco Pais, 1049-001 Lisboa, Portugal.}\\
$^{2}${\small\em Department of Physics, Karlstad University,}\\
{\small\em S-65188 Karlstad, Sweden.}\\
$^{3}${\small\em Department of Applied Mathematics, University of Waterloo,}\\
{\small\em Waterloo, ON, N2L 3G1, Canada.}}


\date{}
\maketitle

\begin{abstract}
The observational success and simplicity of the $\Lambda$CDM model,
and the explicit analytic perturbations thereof, set the standard for any alternative
cosmology. It therefore serves as a comparison ground and as a test case
for methods which can be extended and applied to other cosmological models.
In this paper we introduce dynamical systems and methods to describe linear
scalar and tensor perturbations of the $\Lambda$CDM model, which
serve as pedagogical examples that show the global illustrative powers of
dynamical systems in the context of cosmological perturbations.
We also study the asymptotic
properties of the shear and Weyl tensors and discuss the validity of the
perturbations as approximations to the Einstein field equations.
Furthermore, we give a new approximation for the linear growth rate,
$f(z) = \frac{d\ln \delta}{d\ln a} = \Omega^{\frac{6}{11}}_m - \frac{1}{70}(1-\Omega_m)^{\frac{5}{2}}$,
where $z$ is the cosmological redshift, $\Omega_m=\Omega_m(z)$, while $a$ is the background scale
factor, and show that it is much more accurate than the
previous ones in the literature.

\end{abstract}

\section{Introduction}

This paper is the first in a series of papers dealing with cosmological
perturbations by means of dynamical systems formulations and methods.
We will show how a wide variety of increasingly complex problems can be formulated
as dynamical systems and how powerful dynamical systems methods can be applied to yield
insights about cosmological perturbations. In particular, this will make it possible
to apply approximation techniques from the theory of dynamical systems and to
obtain illustrative pictures as well as mathematically rigorous results about the
global structure of the various models solution spaces, thereby also providing a
context for especially physically interesting solutions. Our aim is thus to provide
a useful complement to traditional approaches to cosmological perturbations.
Step by step we will introduce increasingly sophisticated models and methods.
In this paper we focus on first order scalar and tensor perturbations of the spatially
flat $\Lambda$CDM model with dust and a positive cosmological constant $\Lambda$.
Since this model is mathematically simple and is compatible with a wide range of
observations it is a natural choice as a first example to illustrate the most simple
aspects of our dynamical systems approach to perturbative cosmology. Moreover, 
due to their observational success they provide a comparative testing ground
for any observational contender.

Dynamical systems have been used before to analyze cosmological linear scalar
perturbations in general relativity (GR) with an open Robertson-Walker (RW) geometry as
background, see~\cite{waiell97} and references therein. These models turn out to be somewhat
easier to handle than models with a spatially flat background,\footnote{In contrast to the
spatially flat RW models, ${\cal H}^{-2}= (aH)^{-2}$ is always bounded for the
spatially open models if the energy density is non-negative. This is due to how the
spatial curvature term appears in the background Gauss constraint, which also can be used to
solve for ${\cal H}^{-2}$. Since ${\cal H}^{-2}$ appears in connection with the spatial
derivatives of the perturbed equations, this significantly simplifies the analysis of the
perturbed field equations in the open case.} which is what we
will focus on. However, there has also been some previous work on dynamical systems in this
area in GR, notably~\cite{brupio94} which treated dust and radiation as a single fluid.

The main foundation for standard cosmology is the spatially flat RW geometry, characterized by
a line element that can be written as
\begin{equation}\label{RW}
ds^2 = - dt^2 + a^2\gamma_{ij} dx^i dx^j = a^2\left(- d\eta^2 + \gamma_{ij} dx^i dx^j \right)
= - H^{-2}dN^2 + a^2\gamma_{ij} dx^i dx^j,
\end{equation}
where $a(t)$ is the background scale factor, $\gamma_{ij}$ is the flat spatial 3-metric, which in Cartesian
coordinates is given by $\delta_{ij}$, and $H = a^{-1}da/dt$ is the background Hubble variable.
The different time coordinates above are the clock time $t$, the conformal time $\eta$, and
the $e$-fold time
\begin{equation}
N = \ln x, \qquad x = \frac{a}{a_0},
\end{equation}
where $N$ describes the number of background $e$-foldings with respect to some reference
epoch $a=a_0$ at which $x=1$ (a negative $N$ describes the number of $e$-folds before
the reference time). If this reference epoch is the present time, then
\begin{equation}
x = \frac{1}{1+z},
\end{equation}
where $z$ is the cosmological background redshift. Much of the work in cosmological
perturbation theory uses the clock time $t$ or the conformal time $\eta$, but we find
it more convenient to use the $e$-fold time $N$ as the starting point for our work.

The outline of the paper is as follows. In the next section we describe some aspects of the
framework of our new dynamical systems approach to cosmological perturbations. In
Section~\ref{sec:scalar} we consider linear scalar perturbations of the $\Lambda$CDM models
from a dynamical systems perspective. We also present and
discuss previously known analytic
results in the present context. In addition we describe and discuss various
asymptotic approximation methods and give a new more accurate approximation for the
so-called linear growth rate. In Section~\ref{sec:tensor} we consider the linear tensor perturbations
from a dynamical systems perspective. In Section~\ref{sec:shearweyl} we relate the present
work to the full state space of GR and discuss the validity of the perturbations as approximations
to the Einstein field equations. This is done by comparing asymptotic perturbative results
with asymptotic results in GR for the Hubble-normalized comoving shear and Weyl tensor.
The paper is concluded with some final remarks in
Section~\ref{sec:concl}.

\section{Dynamical systems approach to cosmological perturbations}\label{sec:framework}

Our aim is to analyze cosmological perturbations by formulating the
governing equations as \emph{regular dynamical systems on compact state spaces}.
In this paper we will consider first
order scalar and tensor perturbations of the spatially flat $\Lambda$CDM models
with dust and a positive cosmological constant $\Lambda$. However, since similar methods
apply to models whose matter content consists of a perfect fluid with a barotropic
equation of state and with, or without, a cosmological constant it is useful to
point out some common features these models exhibit. The governing equation for scalar
perturbations for this class of models can be a single second order
partial differential equation (e.g. the Bardeen equation) or a system of two
coupled first order partial differential equations (e.g. the  Kodama-Sasaki
equations), depending on the choice of gauge and the choice of variables, see
e.g.~\cite{uggwai18} and references therein. For an arbitrary equation of state
these governing equations will contain the spatial Laplacian ${\bf D}^2$. This
is also the case for the tensor perturbations which obey a linear second
order differential master equation for a single variable.

In order to obtain ordinary differential equations (ODEs) we make a
spatial Fourier decomposition of the perturbation variables, which involves
replacing the perturbation variables by their Fourier coefficients and making
the transition
\begin{equation}\label{Dk}
{\bf D}^2 \rightarrow - k^2,
\end{equation}
where $k$ is the wave number. At this stage if we have a second order
ODE as a governing equation we would replace it by a system of
two coupled first order ODEs by using the first order time derivative as an independent
variable. In this way the Einstein field equations for first order scalar and tensor
perturbations of the models under consideration can
be written as a system of linear ODEs of the form
\begin{subequations}\label{uusys1}
\begin{align}
u_1^\prime &= bu_1 + cu_2,\\
u_2^\prime &= du_1 + eu_2,
\end{align}
\end{subequations}
for two linear perturbation variables $u_1=u_1(N,k^2)$ and $u_2=u_2(N,k^2)$, where a
${}^\prime$ denotes differentiation with respect to the $e$-fold time $N$.
Due to the spatial Fourier decomposition one obtains complex variables, but thanks to that
the differential equations are linear, real and complex parts satisfy the same equations,
and therefore we can without loss of generality consider $u_1$ and $u_2$ to be real,
as are $b(N,k^2), c(N,k^2), d(N,k^2), e(N,k^2)$, whose specific form is
determined by the background model. This is exemplified in detail in
section~\ref{sec:tensor} where tensor perturbations are treated.

The next step toward obtaining a regular dynamical system on a bounded state space
is to introduce polar coordinates $u_1 = r\cos\theta$, $u_2 = r\sin\theta$,
which thanks to the linearity of the equations lead to a decoupling of an equation for $r$
and a reduction of the system of two linear ODEs~\eqref{uusys1}
to one nonlinear ODE for $\theta(N,k^2)$:
\begin{equation}\label{thetadiff}
\begin{split}
\theta^\prime &= d\cos^2\theta + (e-b)\sin\theta\cos\theta - c\sin^2\theta \\
&= \sfrac12[(d-c) + (d+c)\cos2\theta + (e-b)\sin2\theta]. \end{split}
\end{equation}
However, locally it is more convenient to replace $\theta$ with
\begin{equation}\label{ytheta}
y(N,k^2) = y := \tan \theta =  u_2(N,k^2)/u_1(N,k^2),
\end{equation}
which results in a Riccati ODE given by
\begin{equation}\label{ydiff}
y^\prime = d + (e-b)y - cy^2,
\end{equation}
which describes the essential ``reduced'' dynamics of the problem
since, e.g., $u_1$ can be obtained as a quadrature from the
decoupled equation $u_1'=u_1(b+cy)$ once the equation for $y$ has been solved.

To obtain a dynamical system, i.e. a system of first order autonomous ODEs,
that incorporates the dynamics described by the non-autonomous ODE~\eqref{ydiff}
or~\eqref{thetadiff},
we introduce a new dependent variable $T=F(N)$, where $F$ is a
non-negative, bounded  and increasing
function. It follows that $T$ satisfies an autonomous ODE of the form
\begin{equation} \label{G(T)}
T^\prime = G(T), \quad \text{where} \quad G(T)=F'(F^{-1}(T)).
\end{equation}
Note that $G(T)$ is obtained by differentiating $F(N)$ and
then expressing $N$  in terms $T$  using the inverse function $N=F^{-1}(T)$.
This equation is adjoined to~\eqref{ydiff} or~\eqref{thetadiff},
thereby yielding a 2-dimensional dynamical system for $(y,T)$
or $(\theta,T)$.

In practice, however, we have found it convenient to use the relation $N=\ln(x)$
to express $T=F(N)$ as a function
of $x$ rather than $N$. In particular we write $T$ in terms of a
subsidiary function $h(x)$ according to
\begin{equation}\label{hT}
T = \frac{h(x)}{1 + h(x)},
\end{equation}
where $h(x)$ is a non-negative, increasing, explicitly invertible, and
suitably differentiable function which satisfies $h(0)=0$ and $h(x) \rightarrow \infty$
when $x\rightarrow\infty$. This ensures that $T$ is a non-negative,
bounded and increasing function that satisfies $T(0)=0$ and
$\lim_{x\rightarrow \infty} T(x) = 1$. To find the function $G(T)$
in~\eqref{G(T)}, differentiate~\eqref{hT} with respect to $N$ using
$dx/dN=x$ and then express $x$ in terms of $T$ using~\eqref{hT}.

Augmenting the ODE~\eqref{thetadiff} for $\theta$ with the ODE~\eqref{G(T)}
for $T$ yields the dynamical system
\begin{subequations}\label{thetadiff2}
\begin{align}
\theta^\prime &= d\cos^2\theta + (e-b)\sin\theta\cos\theta - c\sin^2\theta,\\
T^\prime &= G(T),
\end{align}
\end{subequations}
where the functions $b,c,d,e$ are expressed as functions of $T$.
The state space for this system is
a finite cylinder in which all orbits (i.e., solution trajectories),
begin at the boundary $T=0$ and end at the boundary $T=1$. We thus
extend the state space to include these boundaries, which we refer to as the
(extended) compactified \emph{state space cylinder} $[0,1]\times S^1$.
Note that $T$ can be regarded as a bounded time variable for which $T=0$
and $T=1$ correspond to $x=0$ and
$x\rightarrow \infty$, respectively.

As mentioned, for local analysis it is more convenient to use
$y$ instead of $\theta$, which results in the system
\begin{subequations}\label{ydiff2}
\begin{align}
y^\prime &= d + (e-b)y - cy^2,\\
T^\prime &= G(T).
\end{align}
\end{subequations}
In this representation traversing the infinite strip defined by
$-\infty<y<\infty$ and $0\leq T \leq 1$ twice corresponds to
making one revolution on the state space cylinder $[0,1]\times S^1$ with the
global coordinates $T$ and $\theta$.\footnote{The reason for traversing the strip
twice is that $y = u_2/u_1 = -u_2/(-u_1)$, i.e., the mapping is two-to-one,
although note that the right hand side of~\eqref{thetadiff} has a periodicity
$\pi$.}

The key to implementing the above procedure is
to make an appropriate choice for the function $h(x)$ that determines
the new dependent variable $T$ through equation~\eqref{hT}.
In a particular application the choice  is motivated by the
form of the coefficient functions $b,c,d,e$ in the initial
system~\eqref{uusys1}. We will use this procedure in the following
sections to describe the scalar and tensor perturbations of the
$\Lambda$CDM model.

\section{Dynamical systems approach to scalar perturbations of $\Lambda$CDM}\label{sec:scalar}

Linear scalar perturbations of a spatially flat RW background geometry can be described by
\begin{equation}\label{metric_exp}
ds^2 = a^2\left(-(1+2\phi) d\eta^2 +  {\bf D}_i B\,d\eta dx^i +
[(1-2\psi)\gamma_{ij}+ 2{\bf D}_i {\bf D}_j C]dx^i dx^j \right),
\end{equation}
where $\phi, B, \psi, C$ are the metric scalar perturbation variables
(see e.g. Uggla and Wainwright (2018)~\cite{uggwai18}.)
The scalar matter perturbations for $\Lambda$CDM are given by the first order fractional
matter density perturbation
\begin{equation}
\delta_m = \frac{{}^{(1)}\!\rho_m}{{}^{(0)}\!\rho_m},
\end{equation}
where the superscripts denote the order of the perturbation. The scalar velocity perturbation ${V}$
is defined by the first order perturbation of the spatial \emph{covariant} 4-velocity components
according to the relation
\begin{equation}
{}^{(1)}\!u_i = a{\bf D}_i{V}.
\end{equation}

Setting $C=0$ fixes the spatial gauge and covers most of the familiar gauges, although it excludes
the synchronous gauge (for a recent work using the synchronous gauge, see e.g.~\cite{grebru17}, and
also~\cite{uggwai19a} and~\cite{uggwai18} for further discussions and references).
The temporal gauge can be fixed in a number of ways, e.g., by setting to zero one of the
variables $B$, $\psi$, $V$, ${\delta_m}$. We use the following terminology and
subscripts to label the gauges:
\begin{itemize}
\item[i)] Poisson (Newtonian, longitudinal) gauge, subscript ${}_{\mathrm p}$,
defined by $B_{\mathrm p} = 0$,
\item[ii)] uniform curvature gauge, subscript ${}_{\mathrm c}$,
defined by $\psi_{\mathrm c} = 0$,
\item[iii)] total matter (comoving) gauge, subscript ${}_{\mathrm v}$,
defined by $V_{\mathrm v} = 0$.
\end{itemize}
As shown in e.g.~\cite{uggwai18} each choice leads to a different system of governing equations.
In particular, for linear perturbations the uniform (flat) curvature
gauge leads to a simple system of two first order partial differential equations,
which form a natural starting point for a dynamical systems analysis.
We obtain this system of  equations from~\cite{uggwai18}, where it is given in the
following form:\footnote{See equations (10), (54a) and (54b) in~\cite{uggwai18}. We have dropped
superscripts ${}^{(1)}$ on the perturbation variables, and have set $\Gamma=0$
since we are considering a barotropic fluid.}
\begin{subequations}\label{ucg_gov1}
\begin{align}
(1+q)\partial_N((1+q)^{-1} {\phi_\mathrm{c}}) &=
-c_s^2{\cal H}^{-2}{\bf D}^2({\cal H}{B}_{\mathrm c}), \\
\partial_N(a^2\,B_{\mathrm c} )&= -a^2 {\cal H}^ {-1}{\phi_\mathrm{c}},
 \end{align}
\end{subequations}
with
\begin{subequations} \label{prop.q}
\begin{equation} \label{deriv_q}
q^\prime = - (1+q)(1 + 3c_s^2 - 2q).
\end{equation}
Here $q$ is the background deceleration parameter, defined by
either of the following forms,
$q= -{\cal H}^\prime/{\cal H} = -(H^\prime/H) - 1$, where ${\cal H} = aH$.
The background Einstein equations relate $q$ to $w$ according to
(\cite{uggwai18}, equation (4)):
\begin{equation} \label{q.alt}
1+q=\sfrac32(1+w).
\end{equation}
\end{subequations}
For the $\Lambda$CDM model we have\footnote
{See, for example,~\cite{uggwai18}, Appendix B.}
\begin{equation} \label{LCDM1}
c_s^2 = 0, \qquad 1+w= \Omega_m.
\end{equation}

As basic perturbation variables for the $\Lambda$CDM model we
choose $(u_1,u_2)=({\cal H}B_{\mathrm c}, -\phi_{\mathrm c})$,
where we have scaled $B_{\mathrm c}$ with $\cal H$ to
obtain a dimensionless quantity, as discussed
in~\cite{uggwai18} and~\cite{uggwai19a}.
We now specialize the governing equations~\eqref{ucg_gov1} to the
$\Lambda$CDM model, using equations~\eqref{prop.q} and~\eqref{LCDM1}.
After expanding the $\partial_N$ derivatives and replacing $\partial_N$ by $'$
we obtain
\begin{subequations}\label{ODEc}
\begin{align}
\phi_\mathrm{c}^\prime &= -3(1 - \Omega_m)\phi_\mathrm{c}, \\
({\cal H}B_\mathrm{c})^\prime &= -\left(1 + \sfrac32\Omega_m\right)({\cal H}B_\mathrm{c}) - \phi_\mathrm{c}.
\end{align}
\end{subequations}
Since the Laplacian ${\bf D}^2$ has dropped out,
there is no need for a spatial Fourier decomposition.
The system~\eqref{ODEc}, which is of the form~\eqref{uusys1},
constitutes the starting point for transforming the governing equations into
a dynamical system.

We now digress to describe the background dynamics of the $\Lambda$CDM model
in order to obtain the $x$-dependence of $\Omega_m$.
The density parameters are given by
\begin{equation} \label{LCDM2}
\Omega_m =\Omega_{m0}x^{-1} \left(\frac{{\cal H}_0}{\cal H}\right)^2, \qquad
\Omega_{\Lambda} =\Omega_{\Lambda 0}x^{2} \left(\frac{{\cal H}_0}{\cal H}\right)^2,
\end{equation}
which implies that
\begin{equation} \label{LCDM3}
\frac{\Omega_{\Lambda}}{\Omega_m}=\lambda_m x^3, \quad \text{where}
\quad \lambda_m:= \frac{\Omega_{\Lambda0}}{\Omega_{m0}} =  \frac{\Lambda}{\rho_{m0}}.
\end{equation}
Since $\Omega_m + \Omega_{\Lambda} = 1$ it follows from~\eqref{LCDM3} that
\begin{equation} \label{LCDM4}
\Omega_m= \frac{1}{1+\lambda_m x^3}.
\end{equation}
%

\subsection{Derivation of the dynamical systems}

\subsubsection*{The metric perturbations in the uniform curvature gauge}

We have shown that perturbations of the $\Lambda$CDM model are described
by the system of ODEs~\eqref{ODEc} which we repeat here:
\begin{subequations}\label{ODEc.2}
\begin{align}
\phi_\mathrm{c}^\prime &= -3(1 - \Omega_m)\phi_\mathrm{c},\label{ODEc.2.1} \\
({\cal H}B_\mathrm{c})^\prime &=
-\left(1 + \sfrac32\Omega_m\right)({\cal H}B_\mathrm{c}) - \phi_\mathrm{c},
\end{align}
\end{subequations}
with variables $(u_1,u_2)=({\cal H}B_\mathrm{c}, -\phi_\mathrm{c})$. This
problem is explicitly solvable, and we will later give the solution. Here, however,
we will use it as a first illustration of our dynamical systems approach.
We therefore introduce $y=u_2/u_1$, as in~\eqref{ytheta}, but give $y$ a subscript, $y_\mathrm{c}$,
as a reminder that we are using the uniform curvature gauge, thus
\begin{equation} \label{yc.defn}
y_\mathrm{c} = \frac{-\phi_\mathrm{c}}{{\cal H}{B}_\mathrm{c}}.
\end{equation}
It follows from~\eqref{ODEc.2} that
\begin{equation} \label{yc.evol}
y_\mathrm{c}^\prime =
y_\mathrm{c}\!\left(\sfrac52 - \sfrac92 (1-\Omega_m) - y_\mathrm{c}\right).
\end{equation}
To complete the process of constructing a dynamical system
we now have to make a choice for the
additional independent variable $T$. It follows from equation~\eqref{LCDM4}
that
\begin{equation}
1-\Omega_m=\frac{\lambda_mx^3}{1+\lambda_mx^3} = \Omega_\Lambda,
\end{equation}
which suggests that we choose\footnote{The second expression is useful for the calculation that follows.}
\begin{equation} \label{choose.T}
T=\frac{\lambda_mx^3}{1+\lambda_mx^3} = 1 - \frac{1}{1+\lambda_mx^3} = \Omega_\Lambda.
\end{equation}
In other words the function $h(x)$ in~\eqref{hT}
is given by  $h(x)=\lambda_mx^3$, which
has the desired properties. Differentiating~\eqref{choose.T} with respect to $N$
yields
\begin{equation} \label{T.evol}
T'=\frac{3\lambda_mx^3}{(1+\lambda_mx^3)^2} = 3T(1-T),
\end{equation}
after expressing $x$ in terms of $T$ using~\eqref{choose.T} again.
This is the desired autonomous ODE for $T$. Note that $0\leq T\leq 1$
and that $T(x)$ is an increasing function, as required.
We finally use that $1-\Omega_m = T$
in equation~\eqref{yc.evol}.
The resulting equation, with~\eqref{T.evol}, comprises the desired
(analytic) dynamical system as follows:
\begin{subequations}\label{ycTdynsys}
\begin{align}
y_\mathrm{c}^\prime &= y_\mathrm{c}\!\left(\sfrac52 - \sfrac92 T - y_\mathrm{c}\right),\\
T^\prime &= 3T(1-T). \label{ycTdynsys.2}
\end{align}
\end{subequations}
Since $y_\mathrm{c}$ is defined by the ratio of two metric coefficients, one orbit
of the above system corresponds to a one-parameter family of solutions.
Furthermore, the solutions for the dynamical system~\eqref{ycTdynsys} correspond
to solutions with all possible (non-zero) values of $\lambda_m$
since $\lambda_m$ is incorporated in the definition of $T$.
In order to solve for $y_\mathrm{c}(N,x^i)$ one has to impose an
initial condition at $N=0$, which corresponds to the initial epoch $a=a_0$,
of the form $y_\mathrm{c}(0,x^i)=f(x^i),$ where $f(x^i)$ is an arbitrary
spatial function. Note that the initial value of $T(N)$, where $T(N)$ was chosen
in order to arrive at a simple form
for~\eqref{ycTdynsys} and is given by~\eqref{choose.T}, is
$T(0)=1-\Omega_{m0}$ where $\Omega_{m0} = \Omega_{m}(0)$ is the initial
value of $\Omega_m(N)$.

Finally we note that the full dynamical system in the uniform curvature gauge consists of the state space vector $({\cal H}B_\mathrm{c},\phi_\mathrm{c},T)$
governed by the equations~\eqref{ODEc.2} and~\eqref{ycTdynsys.2}.
This state space can then be covered by using polar coordinates
${\cal H}B_\mathrm{c} = r_\mathrm{c}\cos\theta_\mathrm{c}, - \phi_\mathrm{c} = r_\mathrm{c}\sin\theta_\mathrm{c}$,
which leads to a decoupling of $r_\mathrm{c}$ from a dynamical system with a reduced state space
vector $(\theta_\mathrm{c},T)$. Locally it is, however, more convenient to use $y_\mathrm{c}(N,x^i)$
and ${\cal H}B_\mathrm{c}(N,x^i)$  as the decoupled variable, which obeys
\begin{equation}\label{HBc}
({\cal H}B_\mathrm{c})^\prime =
\left(y_\mathrm{c} - \sfrac52 + \sfrac32 T\right)({\cal H}B_\mathrm{c}).
\end{equation}
This equation is easily solved as a quadrature once a solution $y_\mathrm{c}(N,x^i),T(N)$ has been obtained,
thereby also yielding the second constant of the motion which together with the constant of the
motion for the reduced system for $(y_\mathrm{c},T)$ characterizes the various scalar perturbations.

%
%

\subsubsection*{The fractional density perturbation in the total matter gauge}

The density contrast $\delta$ is defined by
\begin{equation}\label{lcdmdelta}
\delta = \frac{{}^{(1)}\!\rho_m}{{}^{(0)}\!\rho_m},
\end{equation}
where ${}^{(0)}\!\rho_m$ is the background matter density
and ${}^{(1)}\!\rho_m$ is the linear density perturbation.
The comoving density contrast  $\delta_{\mathrm v}$,
the gauge invariant that equals $\delta$ in the the total
matter/comoving gauge, plays a central role in observational cosmology.
Since $\delta_{\mathrm v}$ satisfies a second order evolution equation
we can apply our method to analyze its behaviour from a dynamical systems
perspective. In the $\Lambda$CDM universe this evolution equation has
the following form when using $e$-fold time $N$:\footnote
{This evolution equation has been given in a general context in
Uggla and Wainwright (2018)~\cite{uggwai18}, equation (71a.b).
When specialized to $\Lambda$CDM and to first order perturbations,
equation (71a) reads ${\cal L}_D \delta_{\mathrm v}=0$, where the
differential operator ${\cal L}_D$ in (71b) reduces to
${\cal L}_D=\partial_N^2+(1-q)\partial_N -(1+q).$ Since
$1+q=\sfrac32(1+w)=\sfrac32 \Omega_m$ we obtain our equation~\eqref{lcdmdelta}.
To avoid confusion we note that the usual density contrast $\delta$ is defined
by normalizing with ${}^{(0)}\!\rho_m$,
while $\bdelta$ in~\cite{uggwai18} is defined by normalizing with
${}^{(0)}\!\rho + {}^{(0)}\!p$. In the $\Lambda$CDM universe these two normalizations
are the same (see~\cite{uggwai18}, Appendix B).   }
\begin{equation}\label{lcdmdeltaeq}
\delta_{\mathrm v}'' + \left(2 - \sfrac32\Omega_m\right)\!\delta_{\mathrm v}' - \sfrac32\Omega_m\delta_{\mathrm v} = 0.
\end{equation}

In order to formulate this differential equation as a dynamical system we use
$(u_1, u_2)=(\delta_{\mathrm v}' ,\delta_{\mathrm v})$ as basic variables,
and define
\begin{equation} \label{y_v}
y_{\mathrm v}=\frac{\delta_{\mathrm v}' }{\delta_{\mathrm v}},
\end{equation}
in accordance with equation~\eqref{ytheta}.

We obtain the ODE for $y_{\mathrm v}$
by differentiating~\eqref{y_v} using~\eqref{lcdmdeltaeq}, and we make
the same choice~\eqref{choose.T} of $T$ as before. This
results in the following dynamical system
\begin{subequations}\label{yvTeqs}
\begin{align}
y_\mathrm{v}' &= \sfrac32(1-T) - \sfrac12\!\left(1 + 3T\right)\!y_\mathrm{v} - y^2_\mathrm{v},\label{ydiffLCDMdeltav}\\
T' &= 3T(1-T).
\end{align}
\end{subequations}
%

This formulation of the evolution equations which
is based on $y_{\mathrm v}=\delta_{\mathrm v}' /\delta_{\mathrm v}$  differs
from the system that is based on
$y_\mathrm{c} = -\phi_\mathrm{c}/{\cal H}{B}_\mathrm{c}$. However, $y_\mathrm{c}$
and $y_\mathrm{v}$ are closely related, as will be shown later.


\subsection{Dynamical systems analysis \label{dynsys}}

In this section we use dynamical systems methods to obtain qualitative
information about the family of all solutions of the perturbed field equations, including
asymptotic descriptions of the solutions at early and late times.

\subsubsection*{The uniform curvature gauge}

When using the uniform curvature gauge the differential equations that define the
reduced dynamical system are given by equations~\eqref{ycTdynsys},
which we here repeat for the reader's convenience:
\begin{subequations} \label{ycTdynsys2}
\begin{align}
y_\mathrm{c}^\prime &=
y_\mathrm{c}\!\left(\sfrac52 - \sfrac92 T - y_\mathrm{c}\right), \label{DE1}\\
T^\prime &= 3T(1-T). \label{DE2}
\end{align}
\end{subequations}
The associated state space is the infinite strip:
\begin{equation} \label{state.space1}
-\infty< y_\mathrm{c} < \infty, \qquad 0\leq T \leq 1,
\end{equation}
with boundaries $T=0$ and $T=1$.
Alternatively we can use the angular variable $\theta_\mathrm{c}$ defined
by $y_\mathrm{c}=\tan \theta_\mathrm{c}$, which yields the reduced regular global
state space which is a finite section of a cylinder $[0,1]\times S^1$, as
discussed in section~\ref{sec:framework}.

The analysis of this dynamical system is straightforward. Equation~\eqref{DE2}
shows that all the fixed points of the system lie on the boundaries
$T=0$ and $T=1$, and that each orbit is past asymptotic to a fixed point
on $T=0$ and future asymptotic to a fixed point
on $T=1$. The local stability of the fixed points can be determined
in the usual way by linearizing the differential equations.

When specialized to  the
$T=0$ and $T=1$ boundaries the differential equation~\eqref{DE1}
results in the following equations:
\begin{subequations}\label{LCDMX01vc}
\begin{align}
\left.y_\mathrm{c}^\prime\right|_{T=0} &= y_\mathrm{c}\!\left(\sfrac52-y_\mathrm{c}\right) , \\
\left.y_\mathrm{c}^\prime\right|_{T=1} &= -y_\mathrm{c}(2+y_\mathrm{c}).\label{LCDMX1c}
\end{align}
\end{subequations}
It follows that on the $T=0$ boundary there are two fixed points
(four in the case of $\theta_\mathrm{c}$,
although they are connected by the discrete symmetry $\theta \rightarrow \theta + \pi$
and thereby given by the two fixed points for $y_\mathrm{c}$):\footnote{The
first subscript on a fixed point $\mathrm{P}$
denotes the value of $T$ while the + (-) denotes the
larger (smaller) value of $y_\mathrm{c}$.}
\begin{subequations}\label{fixedpointsP0}
\begin{alignat}{3}
\mathrm{P_{0+}}\!: \quad y_\mathrm{c} &= \sfrac{5}{2}; &\qquad  \theta_\mathrm{c} &= \arctan{\sfrac{5}{2}} + n\pi,\\
\mathrm{P_{0-}}\!: \quad y_\mathrm{c} &= 0; &\qquad \theta_\mathrm{c} &= n\pi.
\end{alignat}
\end{subequations}
Linearization shows that the fixed point $\mathrm{P_{0+}}$
is a hyperbolic saddle and that a single
orbit originates from $\mathrm{P_{0+}}$.
To obtain an analytical approximation for this special orbit,
we make a series expansion for $y_\mathrm{c}$ in powers of $T$
and use the system~\eqref{ycTdynsys} to
solve for the coefficients,\footnote{This can be mathematically
justified be relating this series expansion
to a so-called Picard expansion, see e.g.~\cite{ugg89} and references therein.}
which leads to
\begin{equation}\label{ycsaddleT0}
\begin{split}
y_\mathrm{c} &= \sfrac{5}{2} - \sfrac{3^2\cdot 5}{2\cdot 11}T-\sfrac{2 \cdot 3^3\cdot 5}{11^2\cdot 17}T^2+\dots \\
             &= \sfrac{5}{2} - \sfrac{3^2\cdot 5}{2\cdot 11}\lambda_m x^3 +
\sfrac{3^2\cdot 5^3\cdot 7}{2\cdot 11^2 \cdot 17}(\lambda_m x^3)^2 + \dots\,,
\end{split}
\end{equation}
for $T$ and $x$ close to zero.
%
%
%
Next, linearization shows that the fixed point $\mathrm{P_{0-}}$
is a hyperbolic source, and hence
a one-parameter family of orbits originates from $\mathrm{P_{0-}}$.
A series expansion in $T$ results in
\begin{equation}\label{ycsourceT0}
\begin{split}
y_\mathrm{c} &= C_0 T^{\frac{5}{6}}(1 - \sfrac{2}{3}T) - \sfrac25C_0^2 T^{\frac{5}{3}} + \dots \\
             &= C_0(\lambda_m x^3)^{\sfrac56}(1 - \sfrac32\lambda_m x^3) - \sfrac25{C}_0^2(\lambda_m x^3)^{\sfrac53} + \dots\, ,
\end{split}
\end{equation}
where $C_0$ is a spatial function that parameterizes the orbits,
depending on the spatial position.

Inserting the series expansion~\eqref{ycsourceT0} into equation~\eqref{HBc} for ${\cal H}B_\mathrm{c}$
shows that ${\cal H}B_\mathrm{c}\rightarrow \infty$ toward the past for all
orbits that originate from $\mathrm{P_{0-}}$ (to leading order it suffices to insert $y_c=0$ into~\eqref{HBc}).
In contrast inserting~\eqref{ycsaddleT0} into~\eqref{HBc}
shows that ${\cal H}B_\mathrm{c}$ is finite asymptotically for the orbit that originates from $\mathrm{P_{0+}}$.
Since ${\cal H}B_\mathrm{c} = -\psi_\mathrm{p}$
(see equation~\eqref{gaugtrans} below), it follows that the orbit that originates from $\mathrm{P_{0+}}$
\emph{is the only orbit along which the perturbations
remain finite into the past} \emph{i.e.} as $T\rightarrow 0$. Below we will identify this orbit
with the so-called growing mode solution toward the future.
Along all other orbits into the past, i.e., solutions that originate from $\mathrm{P_{0-}}$,
(some of) the perturbations increase without bound and hence will not asymptotically approximate
solutions of the full Einstein equations. This is related to the fact that a general solution
of the perturbed Einstein equations for scalar $\Lambda$CDM perturbations is a linear combination
of a so-called \emph{growing mode} and a \emph{decaying mode}, where the latter has the
property that it becomes unbounded into the past.

On the $T=1$ boundary it follows from~\eqref{LCDMX1c}
that the fixed points are given by
\begin{subequations}\label{fixedpointsP1}
\begin{alignat}{3}
\mathrm{P_{1+}}\!: \quad y_\mathrm{c} &= 0; &\qquad \theta_\mathrm{c} &= n\pi ,\\
\mathrm{P_{1-}}\!: \quad y_\mathrm{c} &= -2; &\qquad \theta_\mathrm{c} &= n\pi - \arctan{2} .
\end{alignat}
\end{subequations}
Linearization shows that the fixed point $\mathrm{P_{1-}}$
is a hyperbolic saddle that attracts a
single orbit that is past asymptotic to $T=0$. A series expansion in $1-T$ yields:
\begin{equation}\label{ycsaddleT1}
\begin{split}
y_\mathrm{c} & = -2 + \sfrac{3^2}{5}(1-T) + \sfrac{3^3}{2^4\cdot 5^2}(1-T)^2+\dots \\
             & = -2 + \sfrac{3^2}{5}(\lambda_m x^3)^{-1} -
\sfrac{3^2\cdot 7\cdot 11}{2^4 \cdot 5^2}(\lambda_m x^3)^{-2} + \dots\, .
\end{split}
\end{equation}
Finally, linearization  shows that the fixed point $\mathrm{P_{1+}}$ is a hyperbolic sink,
which implies that a one-parameter family
of orbits is asymptotic to $\mathrm{P_{1+}}$.  A series expansion in $1-T$ results in
\begin{equation}\label{ycsinkT1}
\begin{split}
y_\mathrm{c} & = C_1 (1-T)^{\frac{2}{3}}(1 - \sfrac{5}{6}(1-T)) +
\sfrac{1}{2}C_1^2(1-T)^{\frac{4}{3}} + \dots \\
             & = {C}_1(\lambda_m x^3)^{-\frac23}\left(1 - \sfrac32(\lambda_m x^3)^{-1}\right) + \sfrac12C_1^2(\lambda_m x^3)^{-\frac43} + \dots\, ,
\end{split}
\end{equation}
where $C_1$ is a spatial function that parameterizes the orbits,
depending on the spatial position.

The preceding analysis of the stability of the fixed points enables
one to predict the qualitative form of the state space~\eqref{state.space1}
in figure 1 which shows the fixed points, the special heteroclinic orbits
(i.e. solution trajectories that originate and end at two distinct fixed points),
$\mathrm{P_{0-}}\rightarrow \mathrm{P_{1-}}$,
$\mathrm{P_{0-}}\rightarrow \mathrm{P_{1+}}$ (note that equation~\eqref{DE1} shows
that this latter orbit is the invariant set $y_\mathrm{c}=0$, which in turn corresponds to the
invariant set $\phi_\mathrm{c}=0$ of equation~\eqref{ODEc.2.1}),
and $\mathrm{P_{0+}}\rightarrow \mathrm{P_{1+}}$, together with
some typical orbits of the dynamical system~\eqref{ycTdynsys2}.
The most significant aspect of the state space in figure 1 is the
heteroclinic orbit $\mathrm{P_{0+}} \rightarrow \mathrm{P_{1+}}$, the growing mode solution, which
represents the most physically important solution of the perturbation equations (actually a
one parameter family of solutions), and which thereby is the primary focus in cosmological
perturbation theory. Note also that due to its observational success, any observational contender
must presumably result in a solution trajectory in $y_\mathrm{c}$ and $T$ that is quite similar 
to that of the growing mode solution for the observational redshift range. 

%
\begin{figure}[ht!]
\begin{center}
\subfigure[]{\label{}
\includegraphics[width=0.52\textwidth]{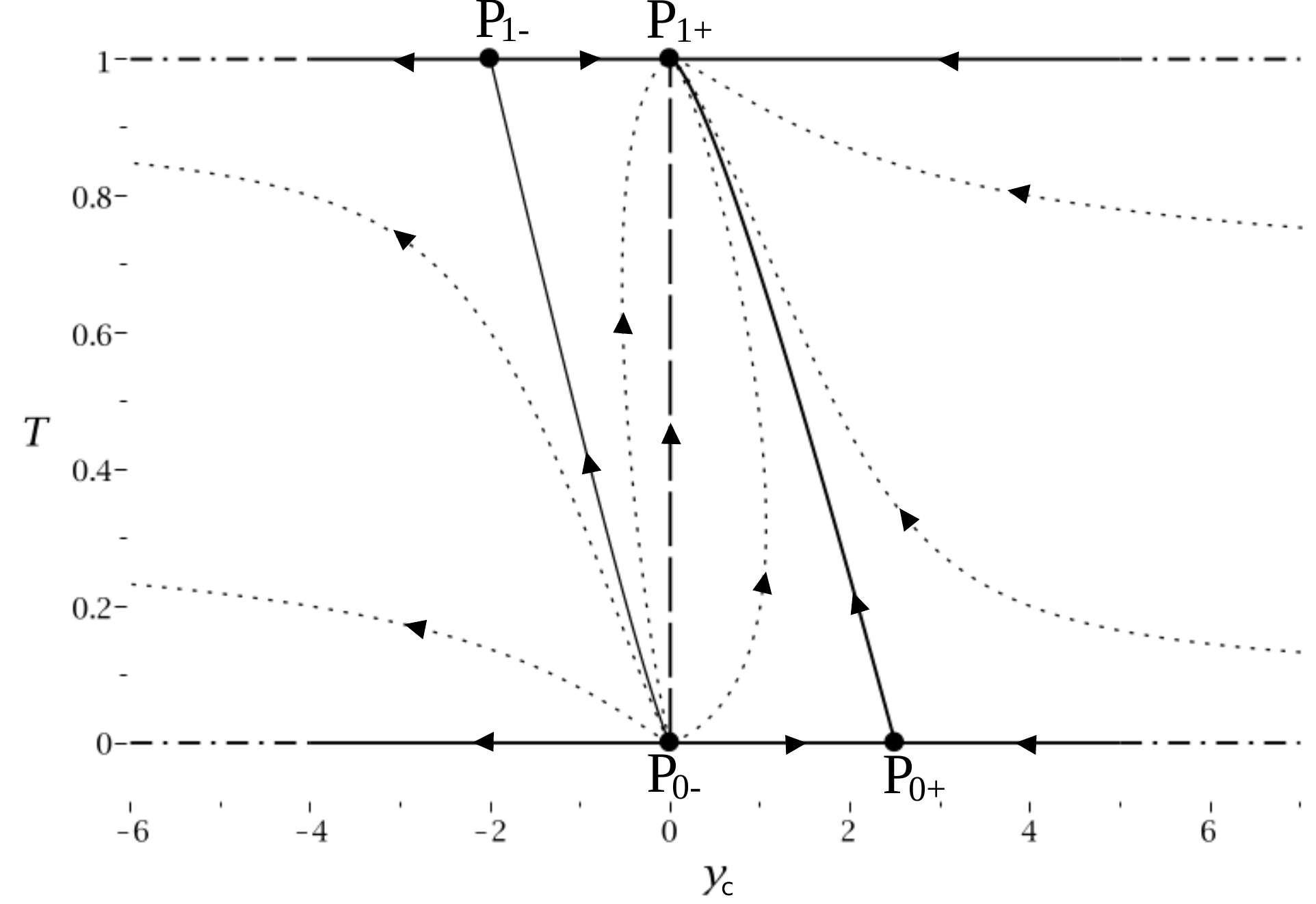}}
\subfigure[]{\label{}
\includegraphics[width=0.45\textwidth]{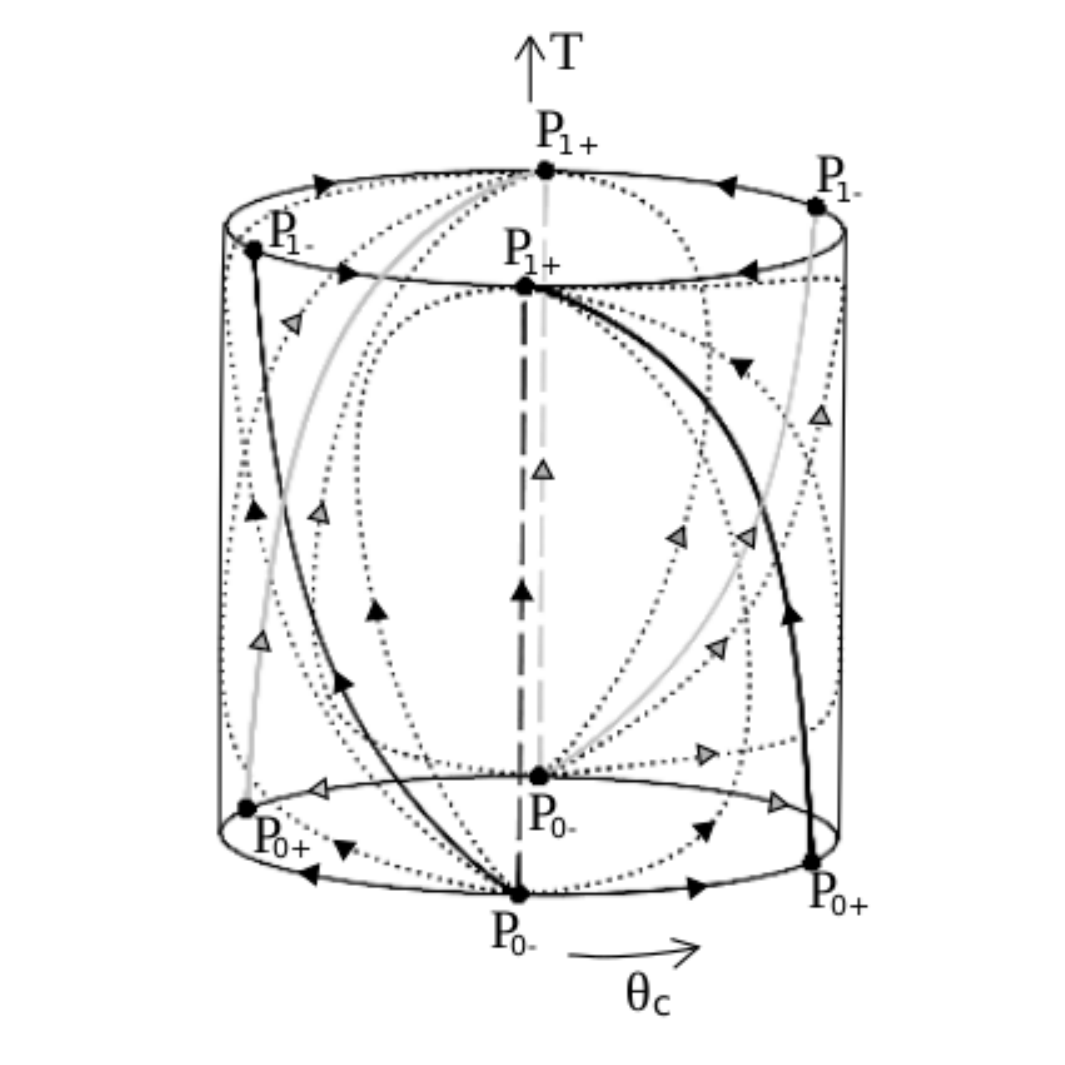}}
\vspace{-0.5cm}
\end{center}
\caption{Solution structure for scalar perturbations of $\Lambda$CDM cosmology in the uniform (flat)
curvature gauge, where $y_\mathrm{c} = \tan\theta_\mathrm{c} = -\phi_\mathrm{c}/{\cal H}B_\mathrm{c}$,  and
$T=1-\Omega_m = \Omega_\Lambda$. The $\mathrm{P_{0+}} \rightarrow \mathrm{P_{1+}}$
solution is the growing mode solution.
}
\label{FigLCDMucg}
\end{figure}

As a final side remark we note the following about the heteroclinic growing mode orbit
$\mathrm{P_{0+}} \rightarrow \mathrm{P_{1+}}$. Since $\mathrm{P_{0+}}$ is a saddle point
\emph{and} $\mathrm{P_{1+}}$ is a local sink, this orbit acts as
an ``attractor solution'' toward the future of nearby
orbits, reminiscent of ``attractor solutions'' in inflationary
cosmology. The latter, however, are associated
with a non-hyperbolic saddle point, whose unstable manifold is a
center  manifold, rather than a hyperbolic saddle point,
which allows the inflationary regime to last
longer than it would if the saddle was hyperbolic, see
e.g.~\cite{alhugg15} for a discussion of attractor
solutions in inflationary cosmology.


\subsubsection*{The comoving density perturbation}

%
%
%
%
%
%
%

The dynamical system based on the comoving density contrast $y_\mathrm{v} = \delta_{\mathrm{v}}'/\delta_{\mathrm{v}}$
is given by~\eqref{yvTeqs}, which we repeat here:
\begin{subequations}\label{yvTeqs2}
\begin{align}
y_\mathrm{v}' &= \sfrac32(1-T) - \sfrac12\!\left(1 + 3T\right)\!y_\mathrm{v} - y^2_\mathrm{v},\label{yv.DE1}\\
T' &= 3T(1-T).
\end{align}
\end{subequations}
The structure of the orbits of this system is very similar to that of the
system~\eqref{ycTdynsys2}, with the fixed points lying
on the boundaries $T=0$ and $T=1$.

When specialized to these boundaries the differential equation~\eqref{yv.DE1}
results in the following equations:
\begin{subequations}\label{LCDMX01v}
\begin{align}
\left.y_\mathrm{v}'\right|_{T=0} &= \sfrac12(1 - y_\mathrm{v})(3 + 2y_\mathrm{v}) , \\
\left.y_\mathrm{v}'\right|_{T=1} &=-y_\mathrm{v} (2 + y_\mathrm{v}).\label{LCDMX1}
\end{align}
\end{subequations}
It follows that on the $T=0$ ($x=0$) boundary the fixed points are given by
\begin{subequations}\label{fixedpointscomov0}
\begin{alignat}{3}
\mathrm{P_{0+}}\!: \quad y_\mathrm{v} &= 1;
&\qquad  \theta_\mathrm{v} &= \sfrac{\pi}{4} + n\pi,\\
\mathrm{P_{0-}}\!: \quad y_\mathrm{v} &= -\sfrac{3}{2};
&\qquad \theta_\mathrm{v} &= -\arctan{\sfrac{3}{2}} + n\pi.
\end{alignat}
\end{subequations}
Linearization shows that the fixed point $\mathrm{P_{0-}}$ is a hyperbolic source,
while the hyperbolic saddle $\mathrm{P}_{0+}$ has
a single orbit originating from it into the interior, which, as we will see,
describes the growing mode solution.
A series expansion for this orbit yields
\begin{equation}
\begin{split}
y_\mathrm{v} &= 1 - \sfrac{2\cdot 3}{11}T -
 \sfrac{2\cdot 3^3\cdot 5}{11^2\cdot 17}T^2 + \dots  \\
             &= 1 - \sfrac{2\cdot 3}{11}\lambda_m x^3 +
              \sfrac{2^2\cdot 3 \cdot 71}{11^2\cdot 17}(\lambda_m x^3)^2 + \dots\, .
\end{split}
\end{equation}
On the $T=1$ boundary the fixed points are given by:
\begin{subequations}\label{fixedpointscomov1}
\begin{alignat}{3}
\mathrm{P_{1+}}\!: \quad y_\mathrm{v} &= 0; &\qquad  \theta_\mathrm{v} &= n\pi,\\
\mathrm{P_{1-}}\!: \quad y_\mathrm{v} &= -2; &\qquad \theta_\mathrm{v} &= -\arctan{2} + n\pi.
\end{alignat}
\end{subequations}
The fixed point $\mathrm{P_{1+}}$ is a hyperbolic sink, while
the fixed point $\mathrm{P_{1-}}$ is a hyperbolic saddle
which attracts a single interior orbit.

There is a one-to-one correspondence as regards fixed points and stability
properties between the dynamical system~\eqref{yvTeqs2} and the dynamical
system~\eqref{ycTdynsys2} for the state space $(y_\mathrm{c},T)$. This correspondence
is reflected in the similarity between the form of the orbits in figure (1a)
and figure (2a).\footnote{Just before submitting the present paper, Basilakos {\it et al.}
published a paper~\cite{basetal19} on the archive with a diagram that corresponds to
figure (2a), but with $\Omega_m$ instead of $T=1-\Omega_m$. They also
gave the explicit solution (their equation (31)) for $y_\mathrm{v}$, which in their
notation was called $U_m$, but instead of using the parameters $C_\pm$ in the next subsection
they used $U_{m0}$ and $\Omega_{m0}$, which are related to $C_\pm$ according to 
$U_{m0} = \frac{C_+T_0^\frac56(1-T_0)^\frac23}{C_+I_0 + C_-} - \frac32(1-T_0)$
where $T_0 = 1 - \Omega_{m0}$ and $I_0 = \frac13\int_0^{T_0}T^{-\frac16}(1-T)^{-\frac13}dT$. 
Our parametrization is adapted to the solution structure so that e.g. the physically 
important growing mode solution is easily obtained by setting $C_-=0$.}
In particular the heteroclinic orbit $P_{0+}\rightarrow P_{1+}$
again represents solutions that only contain the growing mode.
Moreover, since $y_\mathrm{v}$ is positive on this orbit
and $y_\mathrm{v} = \delta_\mathrm{v}'/\delta_\mathrm{v}$
it follows that when $\delta_\mathrm{v}>0$ then $\delta_\mathrm{v}$ is growing
throughout its evolution (hence the name, growing mode),
although $\delta_\mathrm{v}$ approaches a constant value toward
the future since $y_\mathrm{v}=0$ at $\mathrm{P_{1+}}$.\footnote{An
analogue of the orbit $y_{\mathrm c}=0$
in figure (1a) also occurs in figure (2a). It is the heteroclinic
orbit $P_{0-}\rightarrow P_{1+}$ and it is given by
$y_\mathrm{v}=-(3/2)(1-T)$.} Finally, the comoving density contrast $y_\mathrm{v}$
for the growing mode solution is often referred to as the linear growth rate, which we will
denote as $f(z)$ when expressed in terms of the redshift $z$.

\begin{figure}[ht!]
\begin{center}
\subfigure[]{\label{}
\includegraphics[width=0.51\textwidth]{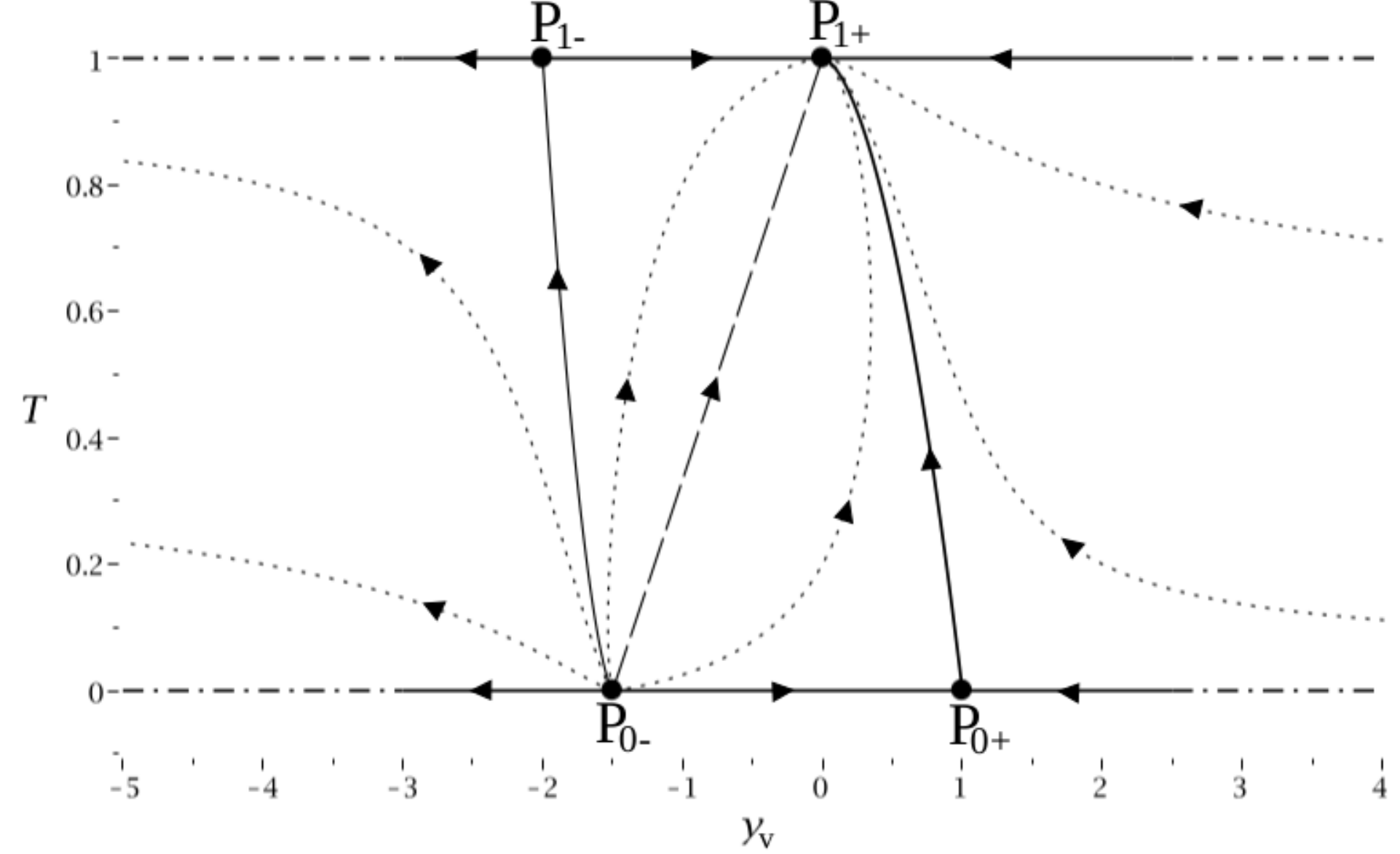}}
\subfigure[]{\label{}
\includegraphics[width=0.45\textwidth]{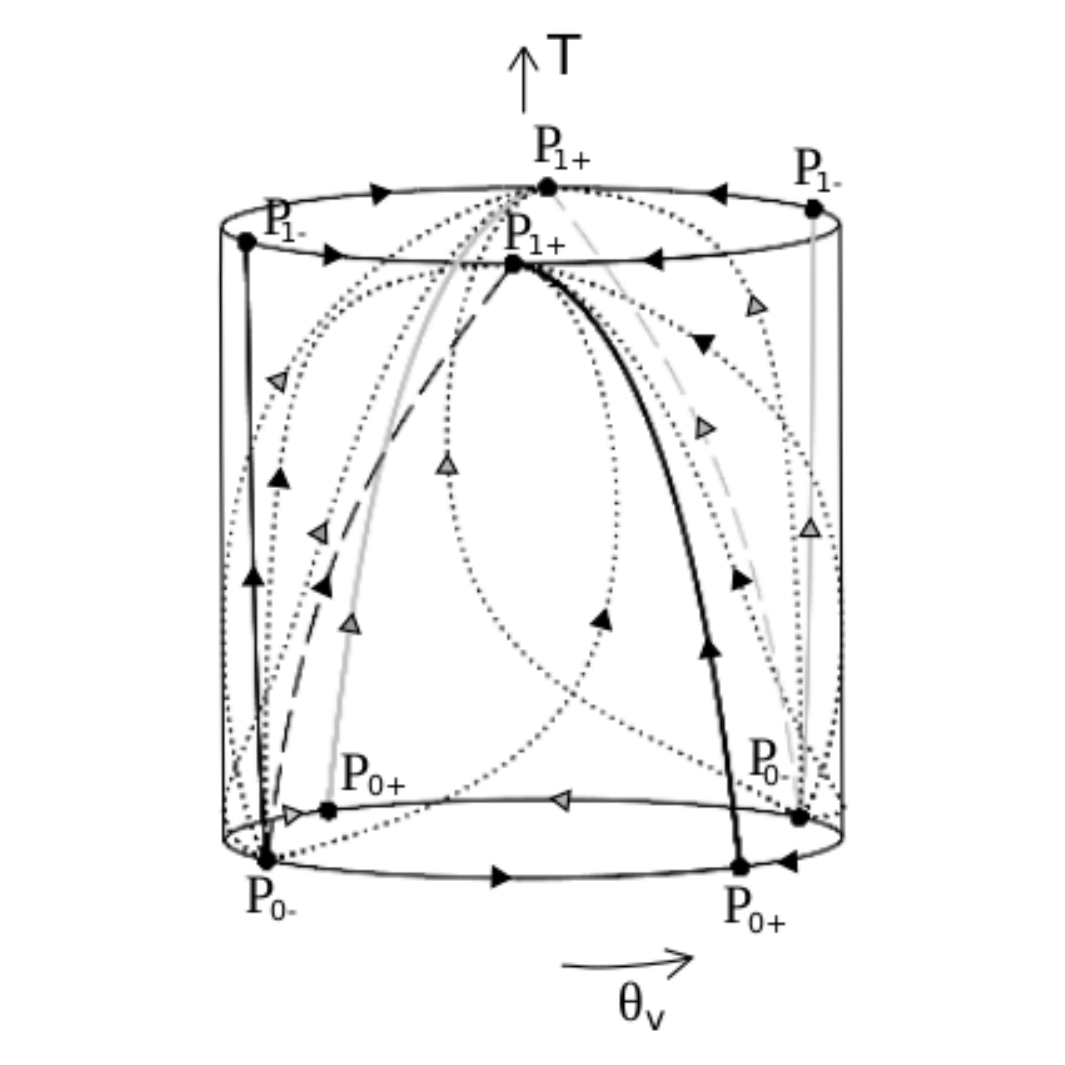}}
\vspace{-0.5cm}
\end{center}
\caption{Global solution structure for $y_\mathrm{v} = \delta_{\mathrm{v}}'/\delta_{\mathrm{v}}$ in $\Lambda$CDM cosmology,
where $T = 1 - \Omega_m = \Omega_\Lambda$ is a monotonically increasing function in the background scale factor.
The $\mathrm{P_{0+}} \rightarrow \mathrm{P_{1+}}$ orbit
describes the growing mode solution.
}
\label{FigLCDMdv}
\end{figure}
%
%

In the above discussion of $y_\mathrm{v}$ we have refrained from giving details about asymptotics,
except for the growing mode solution. The reason for this is that $y_\mathrm{v}$ and
$y_\mathrm{c}$ can be related to each other, as shown in
the next section about explicit solutions. This will also establish the identification of the
growing mode orbits in the $y_\mathrm{c} - T$ and $y_\mathrm{v} - T$ state spaces.

\subsection{Explicit solutions}\label{subsec:explicit}

Here we derive and discuss the explicit solutions for $y_{\mathrm c}$
and $y_{\mathrm v}$ as functions of the time variable $T$.
For the variable $y_{\mathrm c}=-\phi_{\mathrm c}/({\cal H}B_{\mathrm c})$
we begin with
the governing equations~\eqref{ucg_gov1} in the uniform curvature gauge,
which when applied to the $\Lambda$CDM universe simplify to
\begin{subequations}
\begin{align}
\partial_a((1+q)^{-1} {\phi_\mathrm{c}}) &=0, \\
\partial_a(a^2\,B_{\mathrm c} )&= -a {\cal H}^ {-1}{\phi_\mathrm{c}},
\end{align}
\end{subequations}
where $\partial_a$ refers to the partial derivative with respect to the background scale factor $a$.
We can successively solve the equations to obtain
\begin{subequations} \label{phi,B_c,soln}
\begin{align}
\phi_{\mathrm c}&=(1+q) C_{+}, \\
{\cal H}B_{\mathrm c}&= -g(a) C_{+} +
{\cal H}a^{-2}\,C_{ -},
\end{align}
\end{subequations}
where $C_{ +},\,C_{ -} $ are arbitrary spatial functions
and where the \emph{perturbation evolution function} $g(a)$
is given by\footnote{See section 7 in~\cite{uggwai19b} for properties
and a discussion about the perturbation evolution function $g(a)$.}
\begin{equation}
g(a) := \frac{\cal H}{a^2}
\int_0^a \frac{\bar a}{{\cal H}}(1+q)d{\bar a}. \label{def_g}
\end{equation}
In order to obtain $y_{\mathrm c}$ as a function of $T$ we need to
express $\phi_{\mathrm c}$ and ${\cal H}B_{\mathrm c}$ as functions of $T$,
using equations~\eqref{LCDM2}-\eqref{LCDM4},~\eqref{choose.T}
and~\eqref{ycTdynsys.2}.
The results are as follows:\footnote{As intermediate steps we obtain ${\cal H}/a^2=CT^{-5/6}(1-T)^{-1/3}$,
$1+q=\sfrac32 (1-T)$, and \newline $da/a=dT/(3T(1-T))$.
Here $C$ is a constant that does not appear in the final result. The constants
$C_{\pm}$ in~\eqref{phi,B_c,soln} have been redefined in obtaining~\eqref{HBphisol}.  }
\begin{equation}\label{HBphisol}
\phi_\mathrm{c} = -(1-T)C_+,\qquad {\cal H}B_\mathrm{c} =
T^{-\frac56}(1-T)^\frac13(C_+I + C_-),
\end{equation}
where\footnote{$I(T)$ is related to $g$, when expressed in $T$, according to
$g(T)=\frac{3}{2}T^{-\frac{5}{6}}(1-T)^{\frac{1}{3}}I(T)$.}
\begin{equation}\label{ITdefn}
I(T) = \frac13\int^{T}_{0} \bar{T}^{-\frac{1}{6}}(1-\bar{T})^{-\frac{1}{3}}d\bar{T}.
\end{equation}
It follows that the variable $y_{\mathrm{c}}=-\phi_{\mathrm c}/({\cal H}B_{\mathrm c})$
is given by
\begin{equation}\label{ycsol}
y_{\mathrm{c}}= \frac{C_+ T^{\frac{5}{6}}(1-T)^{\frac{2}{3}}}{C_+ I + C_-}.
\end{equation}
The function $I(T)$ is well-defined at the end points,
\begin{equation}
I(0)=0,\qquad I(1)=I_1 =
\frac13\int^{1}_{0}\bar{T}^{-\frac{1}{6}}(1-\bar{T})^{-\frac{1}{3}} d\bar{T},
\end{equation}
and the leading order behaviour of $I(T)$ is given by the following limits:
\begin{equation} \label{I.lim}
\lim_{T\rightarrow 0} \frac{I(T)}{T^{5/6}} = \frac25, \qquad
\lim_{T\rightarrow 1} \frac{I(T)-I_1}{(1-T)^{2/3}} = -\frac12.
\end{equation}
We can now relate the explicit solution~\eqref{ycsol} to the orbits
of the dynamical system in figure 1. It follows from~\eqref{ycsol}
that there are two special values
of $C_{-}$, namely $C_{-}=0$ and $C_{-}=-C_{+}I_1$ that affect
the limit of $y_{\mathrm{c}}$ as $T\rightarrow 0,\,1$, as follows:
\begin{subequations} \label{I.lim}
\begin{alignat}{6}
\lim_{T\rightarrow 0}y_{\mathrm{c}} &= \frac52, &\quad &\text{if} &\quad C_{-} &= 0, &\qquad
\lim_{T\rightarrow 0}y_{\mathrm{c}} &= 0, &\quad &\text{if} &\quad C_{-} &\neq 0,\\
\lim_{T\rightarrow 1}y_{\mathrm{c}} &=-2, &\quad  &\text{if} &\quad C_{-} &= -C_{+}I_1, &\qquad
\lim_{T\rightarrow 1}y_{\mathrm{c}} &=0, &\quad &\text{if} &\quad C_{-} &\neq  -C_{+}I_1.
\end{alignat}
\end{subequations}
Referring to figure 1 we conclude that the orbit with $C_{-}= 0$ is the
growing mode orbit, \emph{i.e.} the orbit that is
past asymptotic to the fixed point $P_{0+}$, while the orbits with $C_{-}\neq0$
are those that are past asymptotic to  the local source $P_{0-}$.
Further, the orbit with $C_{-}= -C_{+}I_1$ is the special orbit
that is future asymptotic to the fixed point $P_{1-}$,
while the orbits with $C_{-}\neq -C_{+}I_1$
are those that are future asymptotic to  the local sink $P_{1+}$.

In section~\ref{dynsys} we derived the leading
terms of a series expansions of $y_{\mathrm{c}}$
for each of the above four classes of orbits. We now
derive a full series for $y_{\mathrm{c}}$ as given by~\eqref{ycsol},
by giving a series expansion for $I(T)$,
first in powers of $T$ and then in powers of $1-T$.
It follows from~\eqref{ITdefn} that (see e.g.~\cite{abrste72})
\begin{equation}\label{ITseries}
I(T) =
\frac{2}{5}T^{\frac{5}{6}}\, {}_{2}F_{1}(\sfrac13,\sfrac56;\sfrac{11}{6};T) =
\frac{2}{5}T^{\frac{5}{6}}\sum^{+\infty}_{n=0} \frac{(\frac{1}{3})_n(\frac{5}{6})_n}{(\frac{11}{6})_n} \frac{T^n}{n!},
\end{equation}
where ${}_{2}F_{1}$ is the Gaussian hypergeometric function,
and $(p)_n$ the Pochhammer symbol, with
$(p)_0=1$, and $(p)_n=p(p+1)...(p+n-1)$,  $n\in\mathbb{N}$.
A truncated version of~\eqref{ITseries} when substituted in~\eqref{ycsol} with $C_{-}=0$
yields\footnote{Making this transition is made somewhat complicated by the fact the series for $I(T)$ is in the
denominator of $y_{\mathrm c}$. Truncate the series after three terms
if $C_{-}=0$ and after one term otherwise.}
the leading term expression~\eqref{ycsaddleT0}, while if $C_{-}\neq0$ we obtain
the leading term expression~\eqref{ycsourceT0} with the arbitrary function
$C_0$ given by $C_0=C_{+}/C_{-}.$

Similarly, we can expand $I(T)$ in powers of $1-T$ obtaining (see e.g.~\cite{abrste72})
\begin{equation} \label{ITseries.2}
\begin{split}
I(T) &= I_1-\frac{1}{2}(1-T)^{\frac{2}{3}}\,{}_{2} F_{1}(\sfrac{1}{6},\sfrac{2}{3};\sfrac{5}{3};1-T)\\
&=I_1-\frac{1}{2}(1-T)^{\frac{2}{3}}\sum^{+\infty}_{n=0}\frac{(\frac{1}{6})_n (\frac{2}{3})_n}{(\frac{5}{3})_n}\frac{(1-T)^n}{n!}.
\end{split}
\end{equation}
A truncated version of~\eqref{ITseries.2} in~\eqref{ycsol}
with $C_{+}I_1 + C_{-} = 0$ yields
the expression~\eqref{ycsaddleT1}, while if $C_{+}I_1 + C_{-} \neq 0$ we obtain
the expression~\eqref{ycsinkT1} with the arbitrary function $C_1$ given by
$C_1=C_{+}/(C_{+}I_1 + C_{-}).$

We can also use the solution for $y_{\mathrm{c}}$ to obtain
an explicit expression for $y_\mathrm{v}$, as follows.
The GR Poisson equation and the conservation of momentum equation
are given by
\begin{subequations}
\begin{align}
\delta_{\mathrm{v}} &= (1+q)^{-1}{\cal H}^{-2} {\bf D}^2\psi_{\mathrm p},\\
\delta_{\mathrm{v}}^\prime &= - {\cal H}^{-2}{\bf D}^2({\cal H}V_\mathrm{p}) ,
\end{align}
\end{subequations}
respectively, see~\cite{uggwai18}.
In addition we have the relations\footnote{The first two are standard change of gauge formulas,
see e.g.~\cite{uggwai19a}, section 3, and the third is the velocity constraint in the uniform
curvature gauge, see ~\cite{uggwai18}, equation (54c).}
\begin{equation}\label{gaugtrans}
\psi_\mathrm{p}=-{\cal H}B_\mathrm{c},
\qquad{\cal H}V_\mathrm{p} ={\cal H}V_\mathrm{c} -{\cal H}B_\mathrm{c},\qquad
 \phi_\mathrm{c} = -(1+q){\cal H}V_\mathrm{c},
\end{equation}
which lead to the following result:
\begin{equation}\label{yv}
y_\mathrm{v} =  \frac{\delta_{\mathrm{v}}^\prime}{\delta_{\mathrm{v}}}=
\frac{{\bf D}^2(-\phi_\mathrm{c})}{{\bf D}^2({\cal H}B_\mathrm{c})} - (1 + q).
\end{equation}
We substitute the explicit solution~\eqref{HBphisol}
into~\eqref{yv} to obtain
\begin{equation}\label{explicityv}
y_\mathrm{v}(N,x^i) = \frac{({\bf D}^2C_+)
T^{\frac{5}{6}}(1-T)^{\frac{2}{3}}}{({\bf D}^2C_+) I + ({\bf D}^2C_-)}   - \frac32(1-T),
\end{equation}
which for the growing mode ($C_-=0$) reduces to
\begin{equation}\label{explicityvgrow}
y_\mathrm{v} = T^{\frac{5}{6}}(1-T)^{\frac{2}{3}}I^{-1} - \frac32(1-T).
\end{equation}
It follows from~\eqref{ycsol} with $C_-=0$ that
for this special orbit we have the simple relation
\begin{equation}\label{yvyc}
y_{\mathrm v} = y_{\mathrm c} - \frac32(1-T).
\end{equation}
This relation then establishes that the heteroclinic orbit $P_{0+}\rightarrow P_{1+}$ in both systems represents
the same solution, i.e., the growing mode solution.
Making a Fourier decomposition of ${\cal H}B_\mathrm{c}$ and $\phi_c$ results in that equation~\eqref{yv}
also reduces to~\eqref{yvyc}. This explains why making the above variable transformation transforms
the system~\eqref{yvTeqs} to the system~\eqref{ycTdynsys2}.

We conclude by remarking that the explicit solutions makes it possible to give an analytic
example of how the growing mode solution acts as an ``attractor solution'' by defining
\begin{equation}
\delta y_{\mathrm{c}} = y_{\mathrm{c}} - y_{\mathrm{c}}[G] = - \frac{C_- T^{\frac{5}{6}}(1-T)^{\frac{2}{3}}}{I(C_+ I + C_-)},
\end{equation}
as follows from equation~\eqref{ycsol} where $y_{\mathrm{c}}[G]$ stands for the growing mode solution with $C_-=0$.
This is an analytic description of the deviation
of solutions from the growing mode solution, which can be given in terms of the redshift $z$ since $T = T(z)$.
Finally, note that it is possible for the spatial function $C_-$ to be zero at
one, two or three spatial coordinates. This gives rise to a so-called permanent spike, which is an asymptotic spatial
discontinuity on a surface, line, or point, respectively, a situation that can be compared with the features of
the special explicit solutions of the Einstein field equations
given in~\cite{limetal04,lim04}.

\subsection{Approximation methods}\label{subsec:approx}

We now turn to approximation techniques and make comparisons with the exact results.
The motivation for this is to develop increasingly accurate
approximation schemes that work for problems that are \emph{not} explicitly solvable. From
comparisons with the explicit solution we find that the various series expansions obtained by
dynamical systems methods give correct asymptotic approximations for the solution. In the present
context it is of particular interest to obtain, preferably globally, or at least for all observable
redshifts, accurate approximations for the growing mode solution, especially with methods that
can be applied to other problems and dynamical systems.

We can improve the accuracy of the previous truncated series approximations by using them to
derive Pad\'e approximants.\footnote{For a discussion of Pad\'e approximants in a cosmological setting
and further references, see e.g.~\cite{alhugg15}.} For example, for the growing mode solution originating
from $\mathrm{P}_{0+}$ in the $y_\mathrm{v}-T$ state space we get the following Pad\'e approximation:
\begin{equation}
y_\mathrm{v} \approx [1,1]_{y_\mathrm{v}} = \frac{1-\frac{3\cdot 7^2}{11 \cdot 17}T}{1-\frac{3^2\cdot 5}{11 \cdot 17}T} =
\frac{1+\frac{2^3\cdot 5 }{11 \cdot 17}\lambda_m x^3}{1+\frac{2\cdot 71}{11 \cdot 17} \lambda_m x^3}.
\end{equation}

A completely different global approximation for the growing mode solution can be obtained by observing that
it is a slowly varying heteroclinic orbit in
$T=\Omega_\Lambda$, or, equivalently $\Omega_m$, as can be seen from figure~\ref{FigLCDMdv}. More precisely,
it is a trajectory that bends slightly to the right in figure~\ref{FigLCDMdv} with respect to the straight
line $y_\mathrm{v}=1-T$ that goes through the fixed points $\mathrm{P}_{0+}$ and $\mathrm{P}_{1+}$.
This motivates the following variable transformation from $y_\mathrm{v}$ to a new function
$\gamma(T)$, defined by 
\begin{equation}
y_\mathrm{v} = (1-T)^{\gamma(T)},
\end{equation}
where $\gamma(T)$ is called the \emph{growth (rate) index} function (for a historical background, see the
discussion below). Expressing equation~\eqref{yvTeqs} as a first order differential equation for $\gamma$
results in
\begin{equation}
3T (1-T)\ln{(1-T)}\frac{d\gamma}{dT}-3T\gamma+\frac{1}{2}\left(1+3T\right)+(1-T)^{\gamma}-\frac{3}{2}(1-T)^{1-\gamma} = 0,
\end{equation}
where a power series expansion of $\gamma(T)$ in $T$ yields
\begin{equation}\label{ExpyvGamma}
y_{\mathrm{v}} = (1-T)^{\frac{6}{11}+\frac{3\cdot 5}{11^2\cdot 17}T +\mathcal{O}(T^2)}.
\end{equation}
%

%
%
%
\begin{figure}[ht!]
     \begin{center}
         \subfigure[]{\label{}
\includegraphics[width=0.50\textwidth]{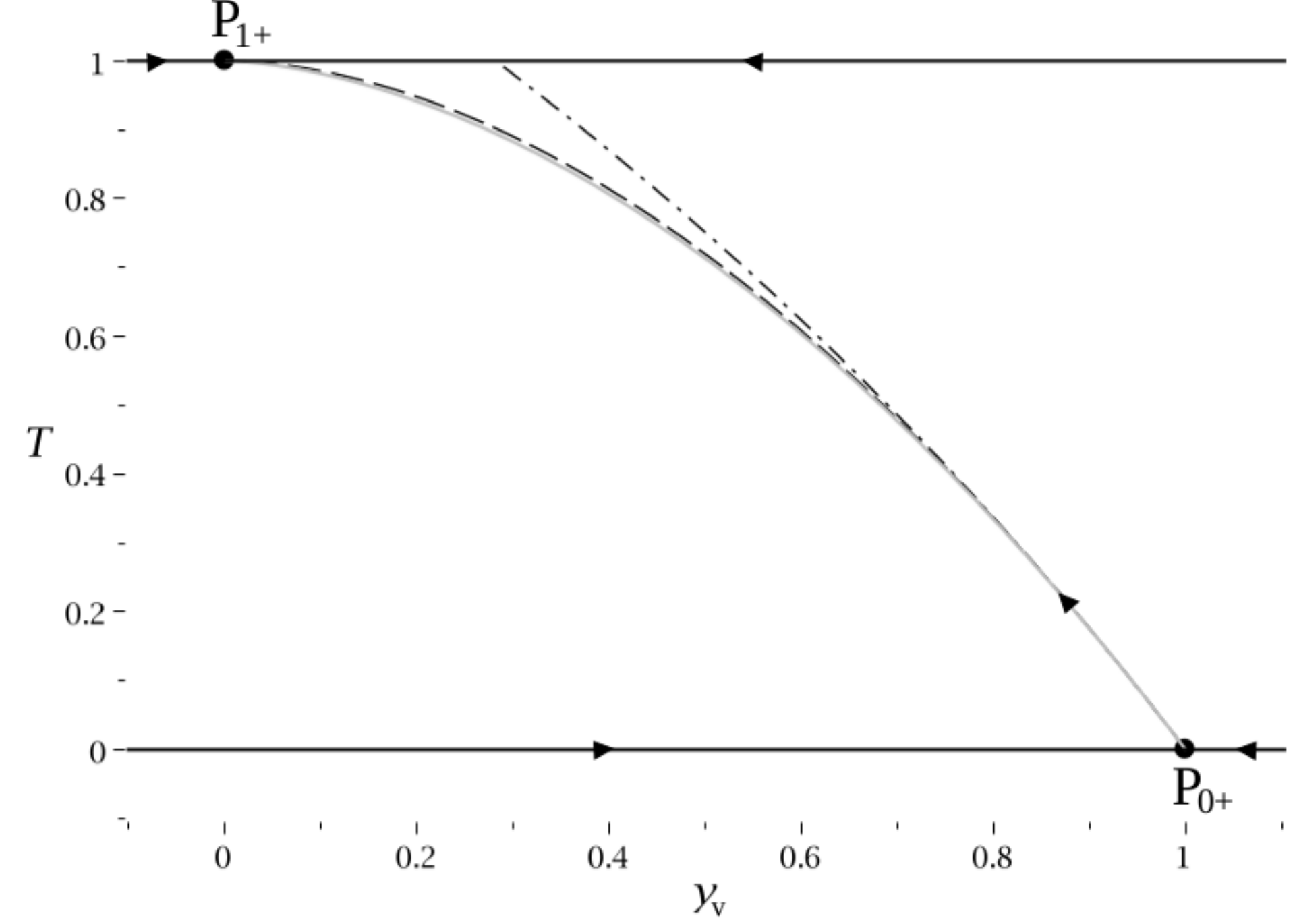}}
         \subfigure[]{\label{}
\includegraphics[width=0.45\textwidth]{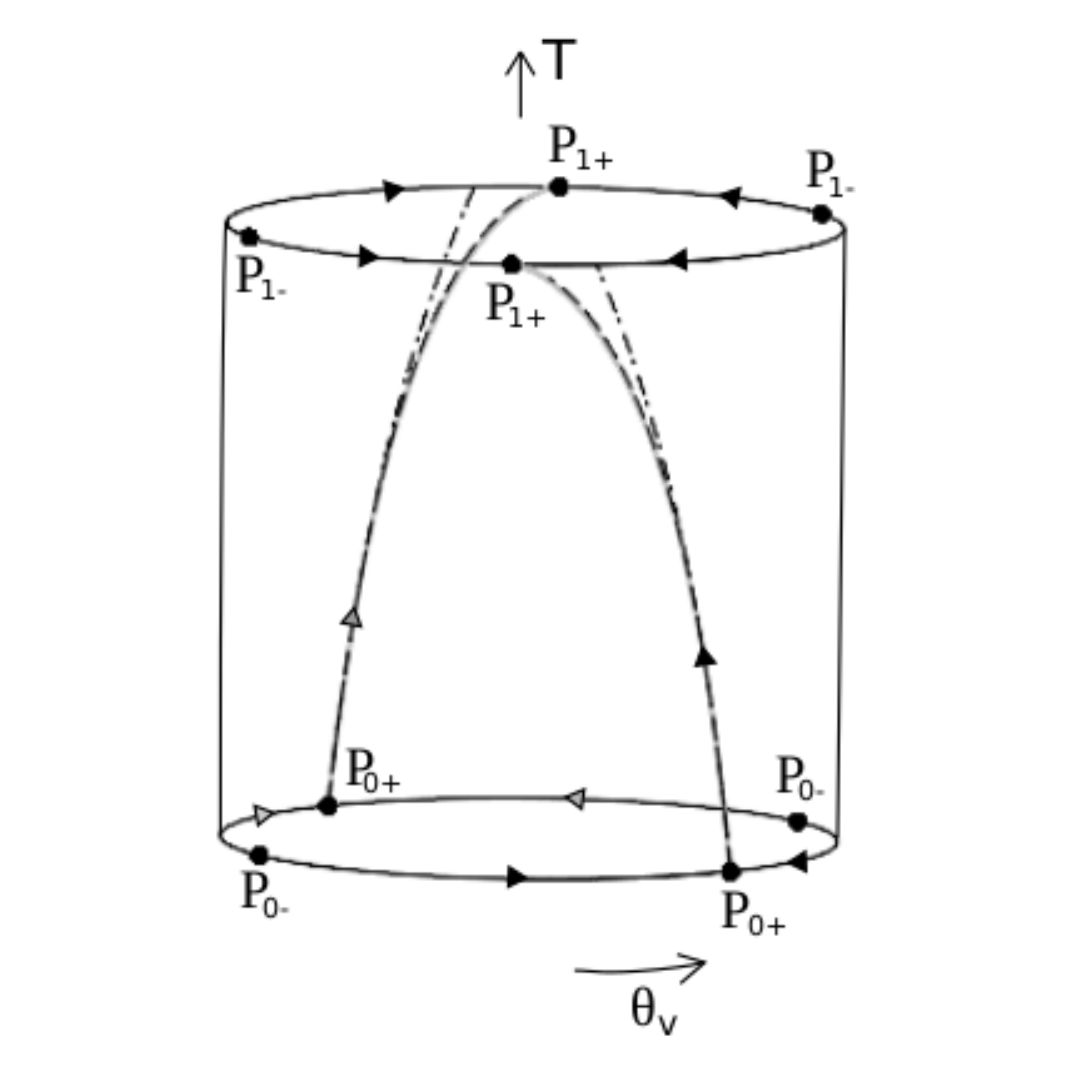}}
         \vspace{-0.5cm}
     \end{center}
     \caption{Explicit growing mode solution (solid grey line),
     the $[1,1]_{y_\mathrm{v}}$ Pad\'e approximant (dashed-dotted),
     $y_\mathrm{v} = (1-T)^{\frac{6}{11}} = \Omega_m^{\frac{6}{11}}$ (dashed).}
     \label{FigLCDApprSS}
\end{figure}
%

%

\begin{figure}[ht!]
     \begin{center}
         \subfigure[The explicit growing mode solution for $\lambda_m=\sfrac73$ (solid) and the one for $\lambda_m=\sfrac13$ (dashed).]{\label{}
             \includegraphics[width=0.48\textwidth]{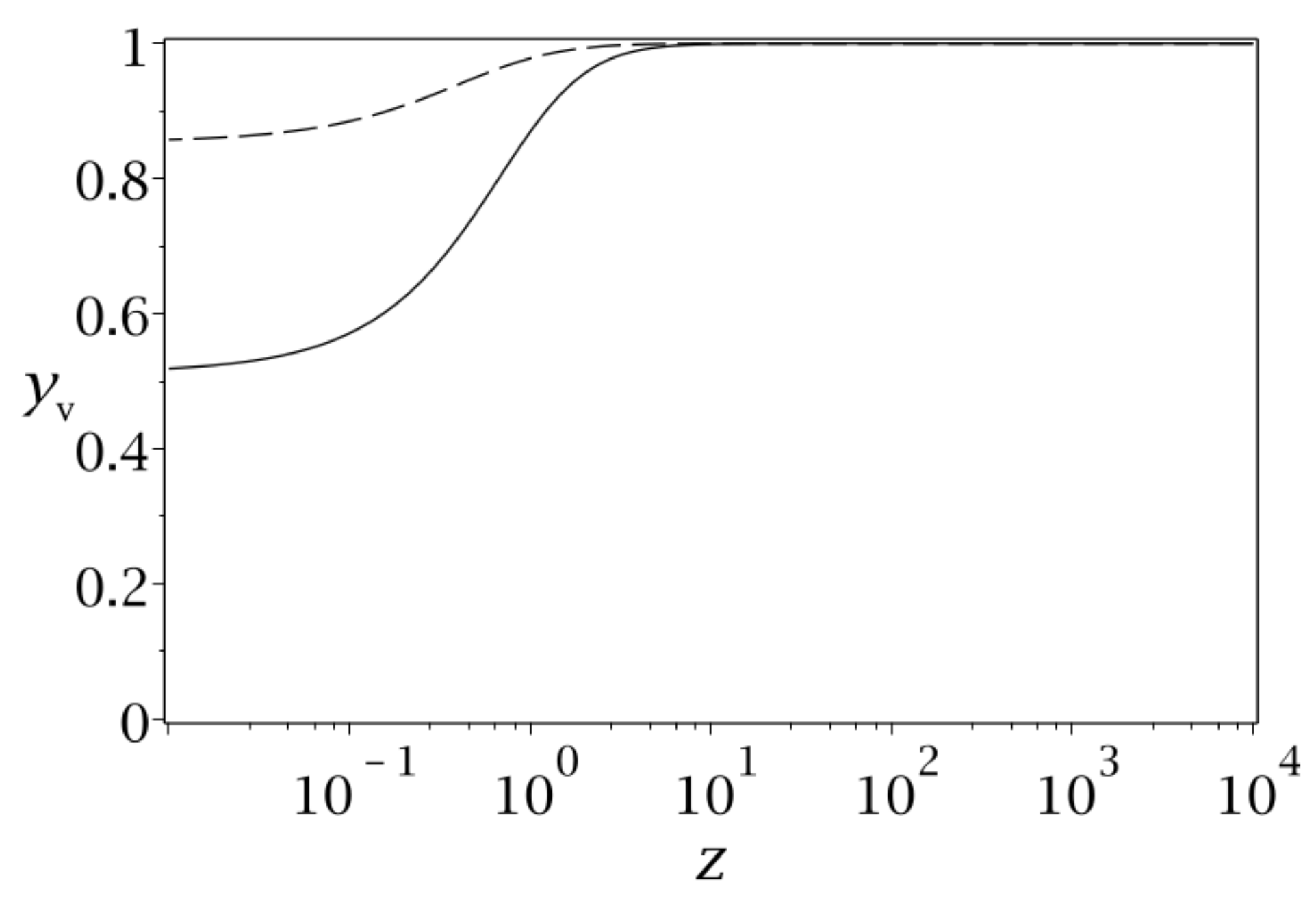}}
         \subfigure[The explicit growing mode solution for $\lambda_m=\sfrac73$ (grey) and its
           ${[1,1]}_{y_\mathrm{v}}$ Pad\'e approximant (dashed-dotted) and the approximation $y_\mathrm{v} = (1-T)^{\frac{6}{11}} = \Omega_m^{\frac{6}{11}}$
           (dashed).]{\label{}
         \includegraphics[width=0.48\textwidth]{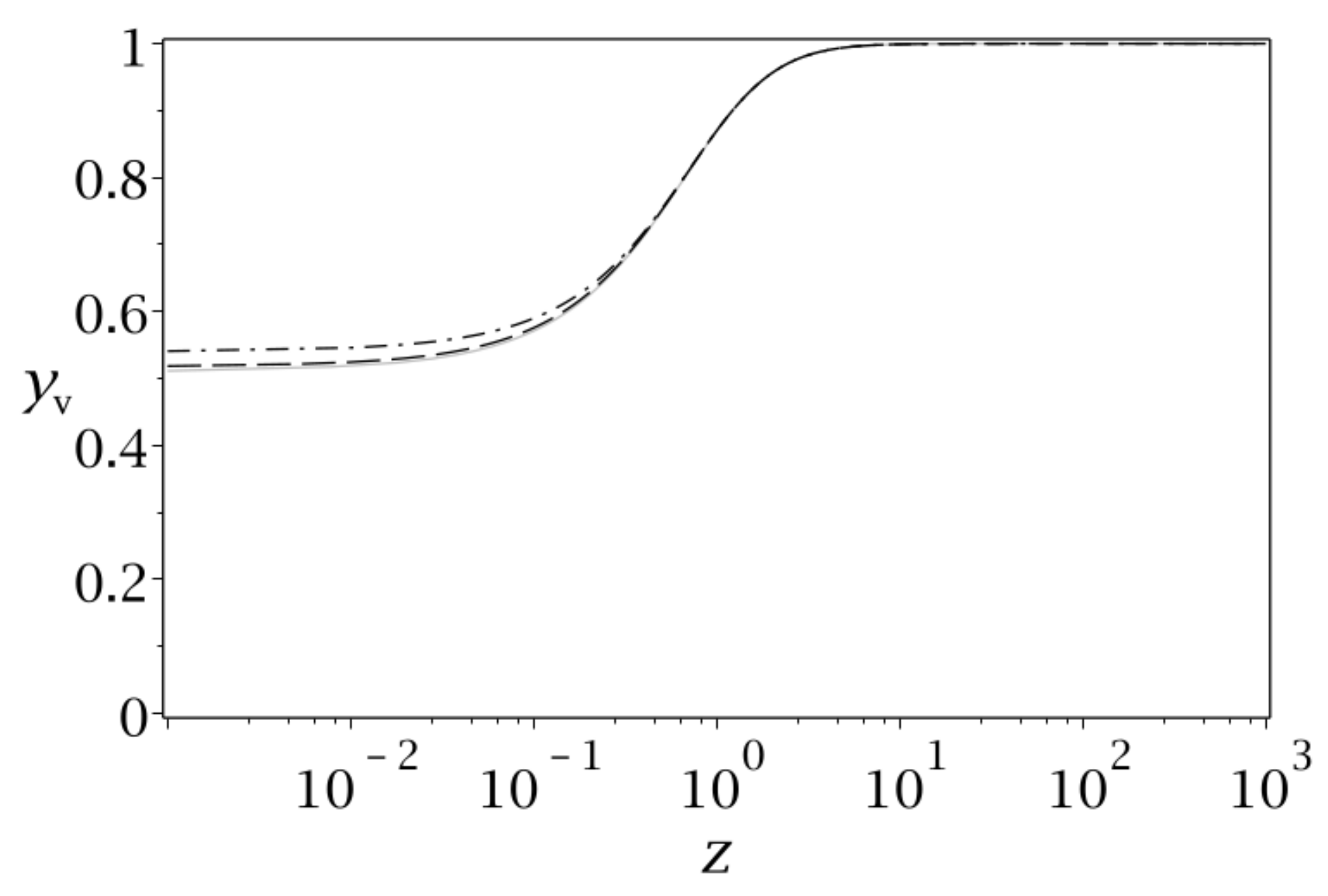}}
         \vspace{-0.5cm}
         \end{center}
         \caption{Plots of the growing mode solution for $y_\mathrm{v}$ as function of the redshift $z=-1+\left(\lambda_m
        \frac{1-T}{T}\right)^{\frac{1}{3}} = -1 + \left(\lambda_m \frac{{\Omega_m}}{1 - {\Omega_m}}\right)^{\frac{1}{3}}$,
        $\lambda_m = \Omega_{\Lambda 0}/\Omega_{m0}$ (where $y_\mathrm{v} = f(z)$ is often referred to as the linear growth rate).}
        \label{FigLCDAppr}
\end{figure}

Background models with different $\Omega_{m0}$ and $\Omega_{\Lambda 0}$, i.e.,
different $\lambda_m = \Omega_{\Lambda 0}/\Omega_{m0}$,
all have the same trajectories in the dynamical systems pictures. However, when e.g. $y_\mathrm{v}$
is plotted against the redshift $z$, a given solution in the dynamical systems picture, e.g. the growing mode
solution, results in a one-parameter set of solutions parameterized by $\lambda_m$ since
\begin{equation}
z = -1 + \left(\lambda_m \frac{1-T}{T}\right)^{\frac{1}{3}} = -1 + \left(\lambda_m \frac{{\Omega_m}}{1 - {\Omega_m}}\right)^{\frac{1}{3}}.
\end{equation}
The growing mode solution together with the $y_\mathrm{v} = [1,1]_{y_\mathrm{v}}$ Pad\'e approximant and the approximation
$y_\mathrm{v} = (1-T)^{\frac{6}{11}} = \Omega_m^{\frac{6}{11}}$  are depicted in a redshift diagram in figure~\ref{FigLCDAppr}
for $\lambda_m=7/3$, i.e., $\Omega_{m0} = 0.3$, $\Omega_{\Lambda 0} = 0.7$ (chosen for simplicity and in agreement
with recent observational data, see e.g.~\cite{planck18}), and
$\lambda_m = 1/3$, i.e., $\Omega_{m0} = 0.75$, $\Omega_{\Lambda 0} = 0.25$.

\begin{figure}[ht!]
     \begin{center}
	\subfigure[Error plots of previously existing approximations]{\label{errora}
		\includegraphics[width=0.48\textwidth]{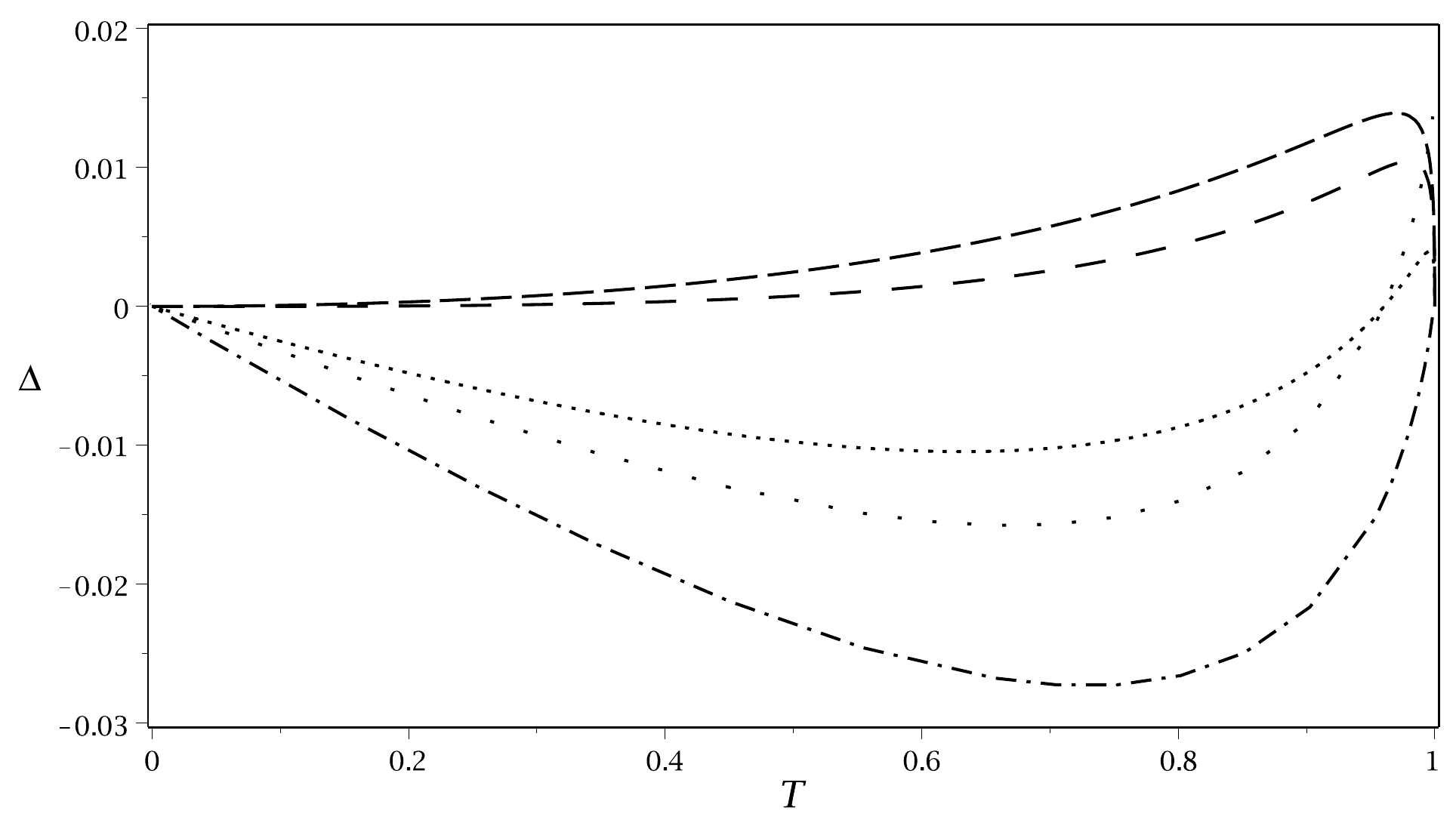}}
\subfigure[Error plot comparing the previous best approximation with the new approximation]{\label{errorb}
	\includegraphics[width=0.48\textwidth]{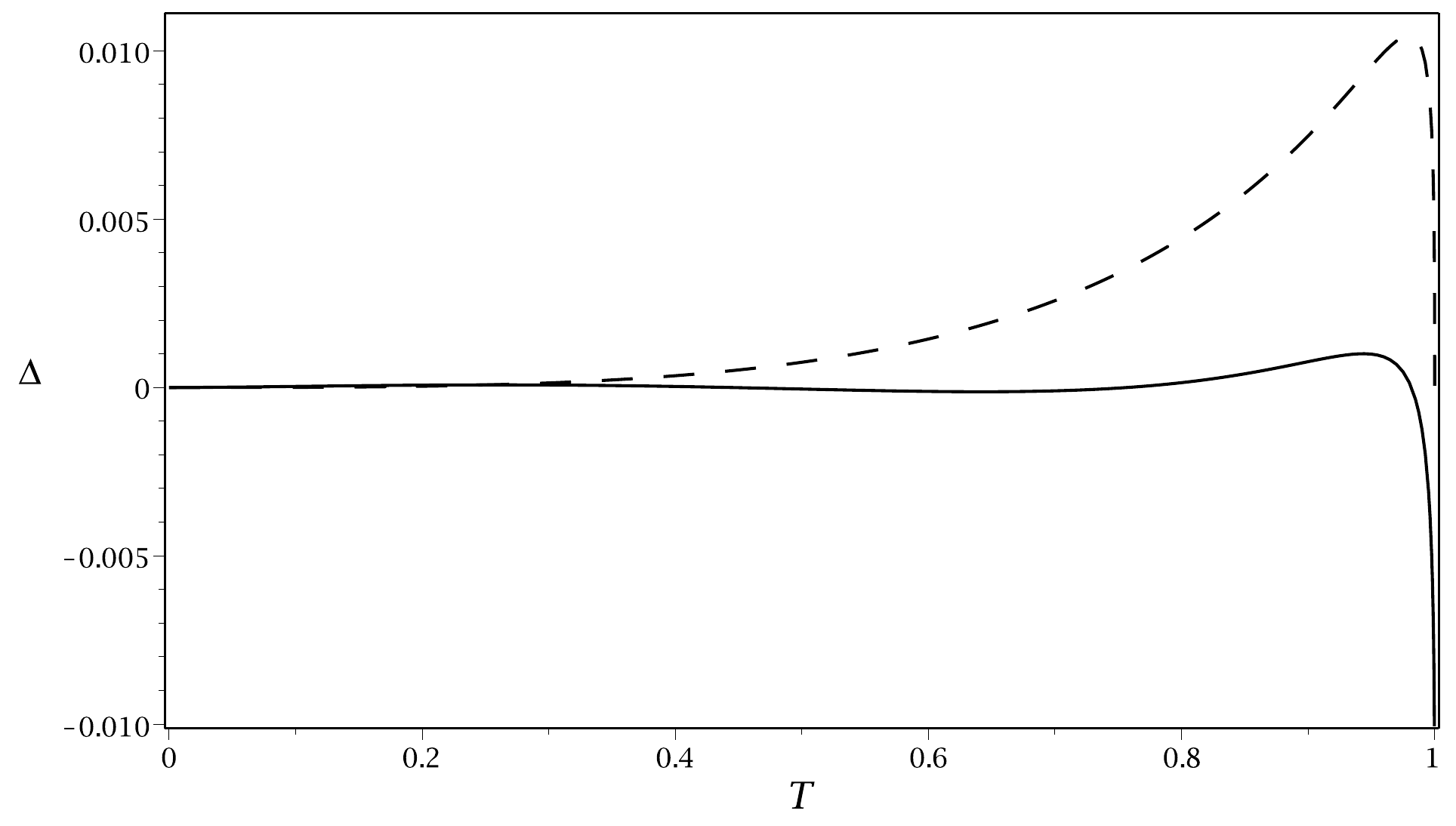}}
	\vspace{-0.5cm}
\end{center}
\caption{Error plots $\Delta = y_\mathrm{v}[\mathrm{Approx.}] - y_\mathrm{v}[\mathrm{Explicit}]$, where
$y_\mathrm{v} = f(z)= \frac{d\ln \delta}{d\ln a}$, for the approximations:
$y_\mathrm{v} = \Omega^{0.6}_m + \frac{1}{70}\left(1-\sfrac{1}{2}\Omega_m(1+\Omega_m)\right)$ (spacedot),
$y_\mathrm{v} = \Omega^{\frac{6}{11}}_m$ (dashed),
$y_\mathrm{v} = \Omega^{\frac{6}{11}+\frac{3\cdot 5}{11^2\cdot 17}(1-\Omega_m)}_m$ (spacedash),
in figure (a) and (b),
$y_\mathrm{v} = \Omega^{\frac{6}{11}}_m - \frac{1}{70}(1-\Omega_m)^{\frac{5}{2}}$ (solid), in figure (b).
Note that $T = 1 - \Omega_m = \Omega_\Lambda$.}
\label{FigLCDNEWAppr}
\end{figure}

Historically the type of analytical approximation given in equation~\eqref{ExpyvGamma} can be traced back
to Peebles~\cite{pee80}, who gave it in the form $y_{\mathrm{v}0}=\Omega^{0.6}_{m0}$ as an approximation
at the present time. For models containing negative curvature $\Omega_k$, a similar approximation
was given by Lightman and Schester~\cite{ligetal90}, with $y_{\mathrm{v}0} = \Omega^{\frac{4}{7}}_{m0}$,
see also~\cite{caretal92}. Lahav {\it et al.}~\cite{lahetal91} realized that this type of
approximations could be extended to general redshift $z$, and they further refined it according to
\begin{equation}\label{Lahav}
y_\mathrm{v}(z) = \Omega^{0.6}_m + \frac{1}{70}\left(1-\sfrac{1}{2}\Omega_m(1+\Omega_m)\right),
\end{equation}
where
\begin{equation}
\Omega_m = \frac{\Omega_{m0}(1+z)^3}{\Omega_{\Lambda 0} + \Omega_{m0}(1+z)^3}, \qquad \Omega_{\Lambda 0} + \Omega_{m0} = 1.
\end{equation}
Later Wang and Steinhardt~\cite{wanste98}, see also references therein,
clarified that the values $0.6$ and $\frac{4}{7}$ are approximations to $\frac{6}{11}$, obtained by the series
expansion given in~\eqref{ExpyvGamma}; see also~\cite{tsuetal13} for the correct next order term in the
exponent for $\Lambda$CDM, which is given in equation~\eqref{ExpyvGamma}.
The approximations in equation~\eqref{Lahav}, $y_{\mathrm{v}} = (1-T)^{\frac{6}{11}} = \Omega_m^{\frac{6}{11}}$,
and $y_{\mathrm{v}} = \Omega_m^{\frac{6}{11}+\frac{3\cdot 5}{11^2\cdot 17}(1 - \Omega_m)}$
are quite good when compared with the explicit solution~\eqref{explicityv}, as shown by plotting the errors
$\Delta = y_\mathrm{v}[\mathrm{Approx.}] - y_\mathrm{v}[\mathrm{Explicit}]$, i.e.
the difference between an approximation and the explicit solution~\eqref{explicityv},
in figure~\ref{errora}. For another discussion about approximations, see~\cite{ham01}.
Let us now introduce the following simple correction to
$y_\mathrm{v} =\Omega^{\frac{6}{11}}_m$, which compensates for the errors for the intermediate 
evolution,\footnote{Most of the approximations in this subsection are algorithmic in nature, but $y_\mathrm{v} = \Omega_m^\frac47$,
$y_\mathrm{v} = \Omega_m^{0.6} + \frac{1}{70}\!\left(1 - \frac12\Omega_m(1 + \Omega_m)\right)$ and our new expression
$y_\mathrm{v} = \Omega_m^\frac{6}{11} - \frac{1}{70}(1 - \Omega_m)^\frac52$ are not. In the dynamical
systems setting they correspond to curve fitting, but all aim at approximating the full analytical
growing mode solution of $y_\mathrm{v}$.}
\begin{equation}\label{NewApprox}
y_\mathrm{v} = f = (1-T)^{\frac{6}{11}} - \frac{1}{70}T^{\frac{5}{2}} =
\Omega^{\frac{6}{11}}_m - \frac{1}{70}(1-\Omega_m)^{\frac{5}{2}},
\end{equation}
where the range of $z$ is determined by $T< T_0 = \Omega_{\Lambda 0}\approx 0.7$.
As seen in figure~\ref{errorb}, equation~\eqref{NewApprox} is an approximation which is more
accurate than
$y_{\mathrm{v}} = \Omega_m^{\frac{6}{11}+\frac{3\cdot 5}{11^2\cdot 17}(1 - \Omega_m)}$
by several orders of magnitude for all observational $z$. It is possible to improve the accuracy
further, but we have not been able to do so significantly with an approximation that is as simple
or simpler than the present one. It should be noted that all approximations are quite good for
large $z$, i.e., small $T$; the differences reside in values for $z$ that are relevant for large
scale structure formation at comparatively late times.
%

\section{Tensor perturbations}\label{sec:tensor}

Linear tensor perturbations of the spatially flat RW background
geometry are characterized by a perturbed metric of the form
\begin{equation}\label{metric_tens}
ds^2 = a^2\left(-d\eta^2 + (\gamma_{ij} + h_{ij})dx^i dx^j \right),
\end{equation}
where $h_{ij}$ is a gauge invariant that satisfies
$\gamma^{ij}h_{ij}=0$, ${\bf D}^ih_{ij} = 0$.
In the absence of anisotropic stresses the perturbed Einstein equations
assume the following form (\emph{e.g.}  Malik and Wands (2009)~\cite{malwan09},
equation (8.6)):
\begin{equation}
\partial_\eta^2 h_{ij} + 2{\cal H}\partial_\eta h_{ij} - {\bf D}^2h_{ij} = 0.
\end{equation}
This equation is now going to be used to illustrate the general discussion in section~\ref{sec:framework}
in more detail. In order to formulate this equation as a dynamical system we first introduce
$e$-fold time, which yields
\begin{equation}
h_{ij}'' + (2-q)h_{ij}' - {\cal H}^{-2}{\bf D}^2h_{ij} = 0.
\end{equation}
Using spatial Cartesian coordinates we apply the Fourier
transform to this partial differential equation and write the transform
of $h_{ij}$ as a linear combination of time-independent polarization tensors
$e_{ij}^{+}$ and $e_{ij}^{\times}$:\footnote{See, e.g., Weinberg (2008)~\cite{wei08}, page 232 for more detail.}
\begin{equation} \label{h-_ij.Fourier}
h_{ij}(N,x^i)\longrightarrow h^{+}(N,k^2)e_{ij}^{+}+h^{\times}(N,k^2)e_{ij}^{\times},  \qquad
{\bf D}^2 \longrightarrow -k^2,
\end{equation}
where $k$ is the wave number, and $h^{+}, h^{\times}$ are complex-valued
functions. The outcome is that each of the functions $h^{+}, h^{\times}$ satisfy
the following ordinary differential equation:
\begin{equation}
h'' + (2-q)h' - {\cal H}^{-2}k^2h = 0.
\end{equation}
In the rest of this section the complex-valued function $h(N,k^2)$ will
denote either $h^{+}$ or $h^{\times}$ and we will usually not indicate the
dependence on the wave number $k$ explicitly.
We finally specialize this differential equation to the $\Lambda$CDM universe
using~\eqref{q.alt},~\eqref{LCDM1} and,~\eqref{LCDM2} to obtain
\begin{equation}\label{tens}
h'' + \sfrac32(2-\Omega_m)h' + k^2_0 (\lambda_m^\frac13 x)\Omega_m h = 0,
\end{equation}
where we have introduced a scaled wave number according to $k^2_0 =
k^2/(\lambda_m^{1/3}\Omega_{m0}{\cal H}_0^2)$.

The final step in constructing a dynamical system (a system of first order autonomous
differential equations) is to follow the approach used for the density perturbation and
introduce a (real-valued) quotient variable analogous to $y_{\mathrm v}$.
However, since $h$ in~\eqref{tens} is complex and stands for one
of the two functions $h^{+}, h^{\times}$, we must write $h^+=h^+_1 +ih^+_2$,
and $h^\times=h^\times_1 +ih^\times_2$
where $h^+_1,h^+_2,h^\times_1, h^\times_2$ are four
real-valued functions which
independently satisfy~\eqref{tens}. Then for any one of these
four functions, which we simply denote by $h=h(N,k^2)$, we define
\begin{equation} \label{y_t.defn}
y_\mathrm{t}(N,k^2) = \frac{h^\prime}{h}.
\end{equation}
To complete the process we have to choose a suitable time function $T$.
Considerations of the temporally $x$-dependent functions in the system of perturbative ODEs
in order to obtain a regular dynamical system result in a different $T$ than for
scalar perturbations, namely\footnote{Compare
with~\eqref{hT} and~\eqref{choose.T}.}
\begin{equation}\label{Ttensor}
T = \frac{\lambda_m^\frac13 x}{1 + \lambda_m^\frac13 x}.
\end{equation}
We now calculate $y_\mathrm{t}^\prime$ by differentiating~\eqref{y_t.defn}
and using~\eqref{tens} and $T'$ by differentiating~\eqref{Ttensor}.
After expressing the coefficients in terms of $T$ using~\eqref{LCDM2}
and~\eqref{Ttensor} we obtain
\begin{subequations}\label{tensdyn}
\begin{align}
y_\mathrm{t}^\prime &= -\left(k^2_0F(T) + G(T)y_\mathrm{t} + y_\mathrm{t}^2\right) ,\\
T^\prime &= T(1-T),
\end{align}
where
\begin{align}
F(T) &= \lambda_m^\frac13x \Omega_m =  \frac{T(1-T)^2}{T^3 + (1-T)^3}, \\
G(T) &= \sfrac32(2-\Omega_m) = \frac32\left[\frac{2T^3 + (1-T)^3}{T^3 + (1-T)^3}\right].
\end{align}
\end{subequations}
Thus equation~\eqref{tensdyn} describes a one-parameter family of real-valued
analytic dynamical systems
labelled by the parameter $k_0^2$, which yields the long
wavelength approximation when
$k_0^2=0$. The state space is again the infinite strip defined
by $-\infty<y_\mathrm{t}<\infty,\, 0\leq T\leq 1$, which is to be traversed twice in
order to describe the global state space of $\theta_\mathrm{t}$ and $T$.

The structure of the orbits of this system is very similar to that of the
system~\eqref{ycTdynsys2}, with $T$ a monotonically increasing function
and with the fixed points lying
on the boundaries $T=0$ and $T=1$. We note that the fixed points do not
depend on the arbitrary parameter $k_0^2$, since on the boundaries the function
$F(T)$ equals zero. When specialized to  these
boundaries the differential equation~\eqref{yv.DE1}
results in the following equations:
\begin{subequations}\label{TensLCDMinvB}
\begin{align}
\left.y_\mathrm{t}^\prime\right|_{T=0} &= -y_\mathrm{t} \left(\sfrac32 + y_\mathrm{t}\right)  , \\
\left.y_\mathrm{t}^\prime\right|_{T=1} &= -y_\mathrm{t} \left(3 + y_\mathrm{t}\right) .
\end{align}
\end{subequations}
It follows that on the $T=0$ boundary the fixed points are given by
\begin{subequations}\label{fixedpointstensorP0}
\begin{alignat}{3}
\mathrm{P}_{0+}\!: \quad y_\mathrm{t} &= 0; &\qquad \theta_\mathrm{t} &= n\pi ,\\
\mathrm{P}_{0-}\!: \quad y_\mathrm{t} &= -\sfrac{3}{2}; &\qquad  \theta_\mathrm{t} &= -\arctan{\sfrac{3}{2}} + n\pi .
\end{alignat}
\end{subequations}

Linearization shows that the fixed point $\mathrm{P_{0+}}$
is a hyperbolic saddle and that a single
orbit originates from $\mathrm{P_{0+}}$  and is future
asymptotic to the $T=1$ boundary.
This special orbit is approximated by
\begin{equation}\label{P0+.tensor}
\begin{split}
y_\mathrm{t} &= -\sfrac{2}{5}k^2_0 T - \sfrac{2}{5}k^2_0(1+\sfrac{2^2}{5\cdot 7}k^2_0)T^2 + \dots \\
             &= -\sfrac{2}{5} k^2_0\lambda^{\frac{1}{3}}_m x - \sfrac{2^3}{5^2\cdot 7} (k^2_0\lambda^{\frac{1}{3}}_m x)^2 + \dots\, .
\end{split}
\end{equation}
%
%
%
%
%
Next, linearization shows that the fixed point $\mathrm{P_{0-}}$
is a hyperbolic source, and hence
a one-parameter family of orbits originates from $\mathrm{P_{0-}}$.
A series expansion in $T$ results in
\begin{equation}\label{P0-.tensor}
\begin{split}
y_\mathrm{t} &= -\sfrac32 + 2 k^2_0 T + C_0 T^{\frac{3}{2}}+\dots  \\
             &= -\sfrac32 + 2k_0^2(\lambda^{\frac{1}{3}}_m x) + C_0 (\lambda^{\frac{1}{3}}_m x)^{\frac{3}{2}} \dots\, ,
\end{split}
\end{equation}
where $C_0$ parameterizes the different orbits.

On the $T=1$ boundary the fixed points are as follows:
\begin{subequations}\label{fixedpointstensorP1}
\begin{alignat}{5}
\mathrm{P}_{1+}\!: \quad y_\mathrm{t} &=  0;  &\qquad  \theta_\mathrm{t} &= n\pi ,\\
\mathrm{P}_{1-}\!: \quad y_\mathrm{t} &= -3;  &\qquad  \theta_\mathrm{t} &= -\arctan{3} + n\pi.
\end{alignat}
\end{subequations}
Linearization shows that the fixed point $\mathrm{P_{1-}}$
is a hyperbolic saddle that attracts a
single orbit that is past asymptotic to $T=0$. A series expansion in $1-T$ yields:
\begin{equation}\label{P1-.tensor}
\begin{split}
y_\mathrm{t} &= -3 + \sfrac{k^2_0}{5} (1-T)^2 + (\sfrac34+\sfrac{2k^2_0}{5}) (1-T)^3 + \dots \\
&= -3 + \sfrac{k^2_0}{5}(\lambda^{\frac{1}{3}}_m x)^{-2}+\sfrac{3}{4}(\lambda^{\frac{1}{3}}_m x)^{-3} + \dots\, .
\end{split}
\end{equation}
Finally, linearization  shows that the fixed point $\mathrm{P_{1+}}$ is a hyperbolic sink,
which implies that a one-parameter family
of orbits is asymptotic to $\mathrm{P_{1+}}$.  A series expansion in $1-T$ results in
\begin{equation}\label{P1+.tensor}
\begin{split}
y_\mathrm{t} &= -k^2_0 (1-T)^2 + C_1 (1-T)^3 + \dots  \\
&= -k^2_0 (\lambda^{\frac{1}{3}}_m x)^{-2} +
(2k^2_0+C_1)(\lambda^{\frac{1}{3}}_m x)^{-3} + \dots \, ,
\end{split}
\end{equation}
where $C_1$ parameterizes the different orbits.

\begin{figure}[ht!]
	\begin{center}
		\subfigure[$k_0=1$]{\label{}
			\includegraphics[width=0.50\textwidth]{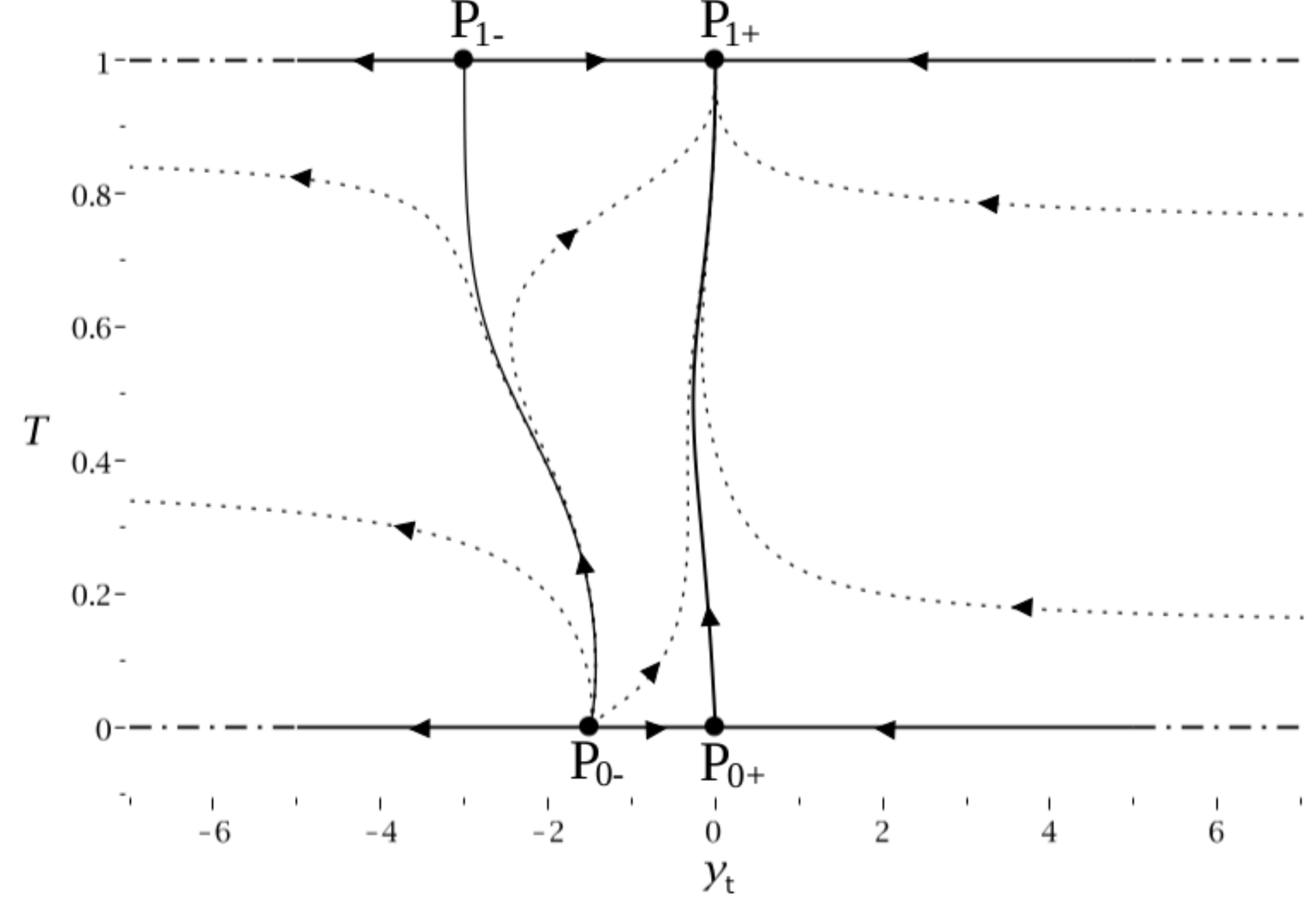}}
		\subfigure[$k_0=1$]{\label{}
			\includegraphics[width=0.45\textwidth]{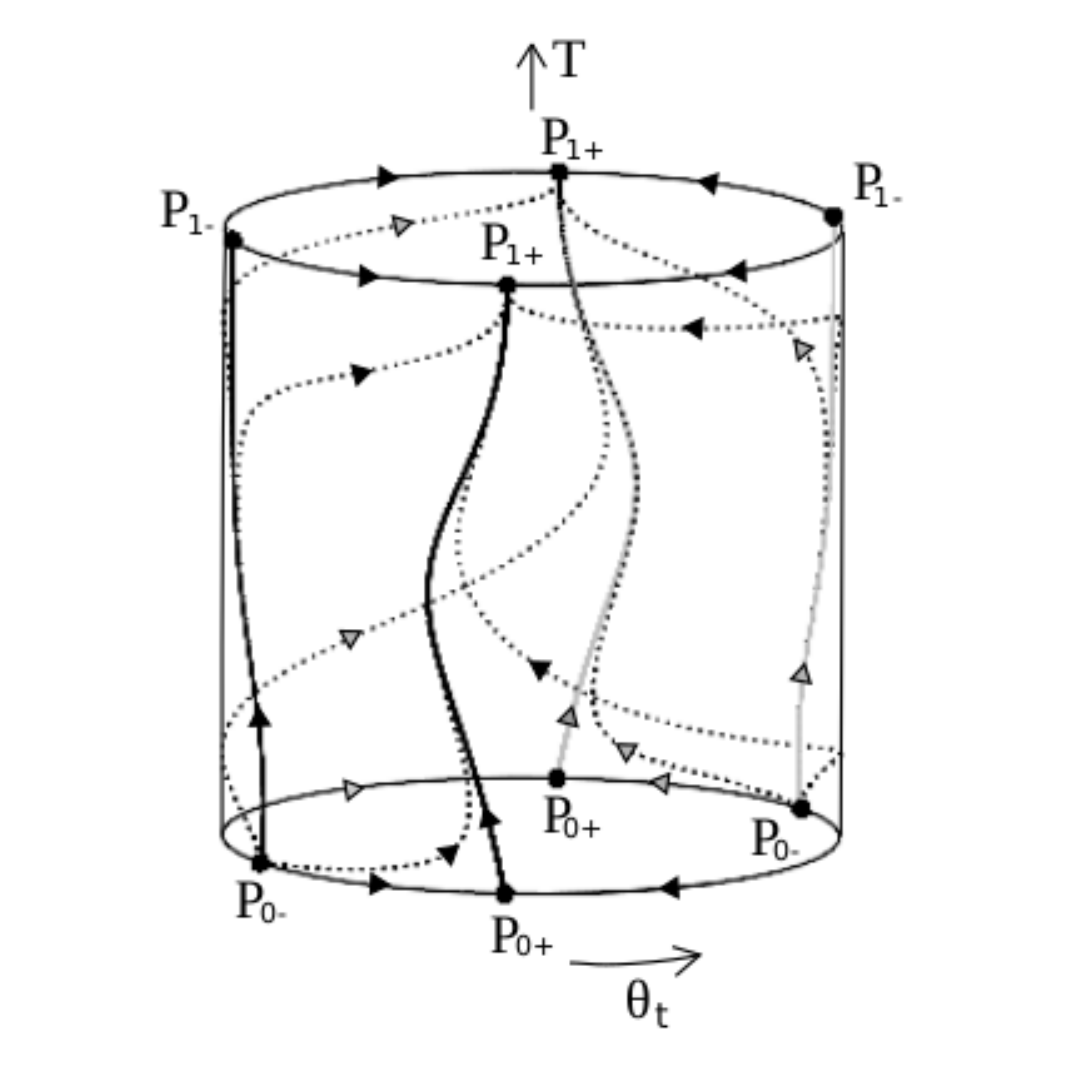}}
        \subfigure[$k_0=5$]{\label{}
			\includegraphics[width=0.45\textwidth]{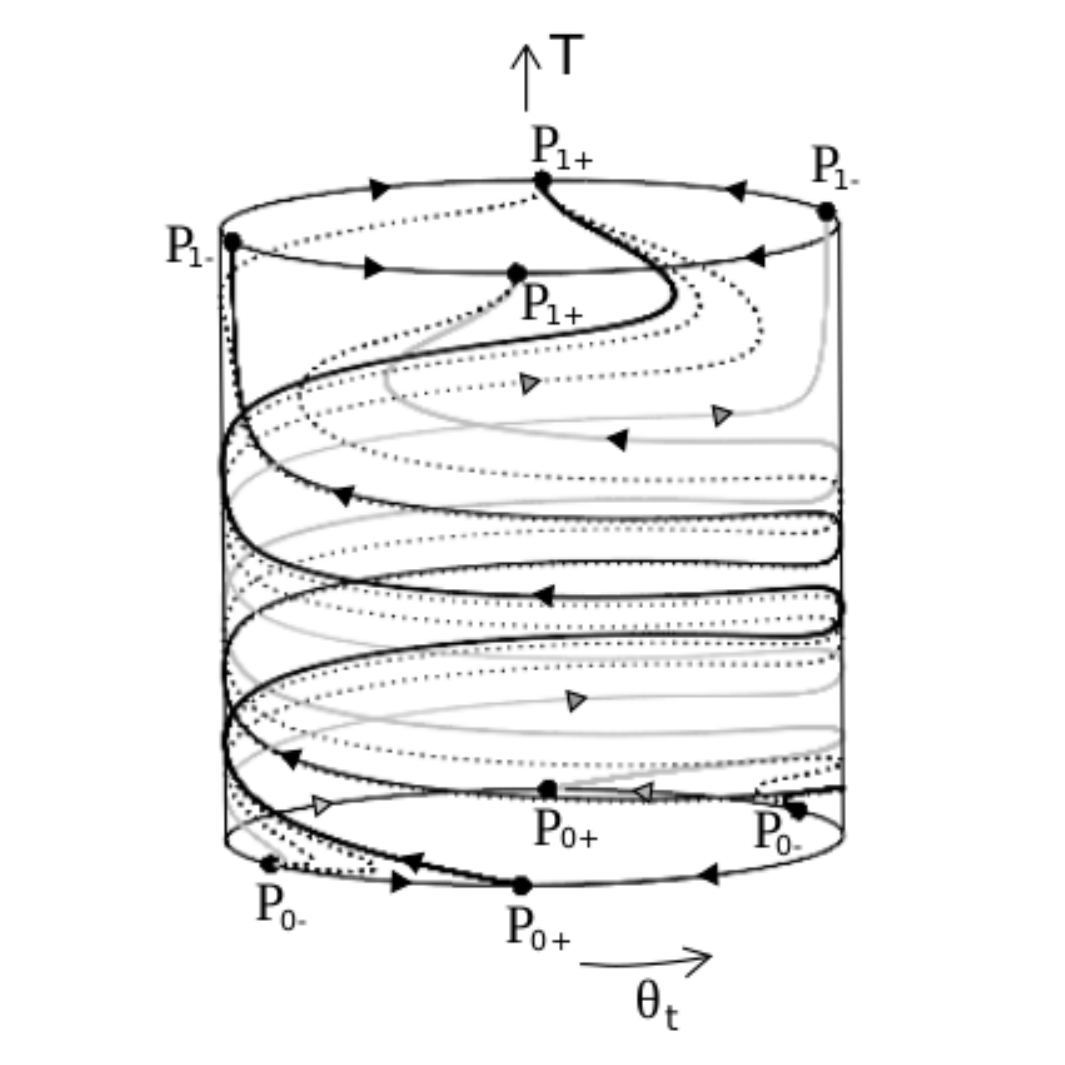}}
        \subfigure[$k_0=7$]{\label{}
			\includegraphics[width=0.45\textwidth]{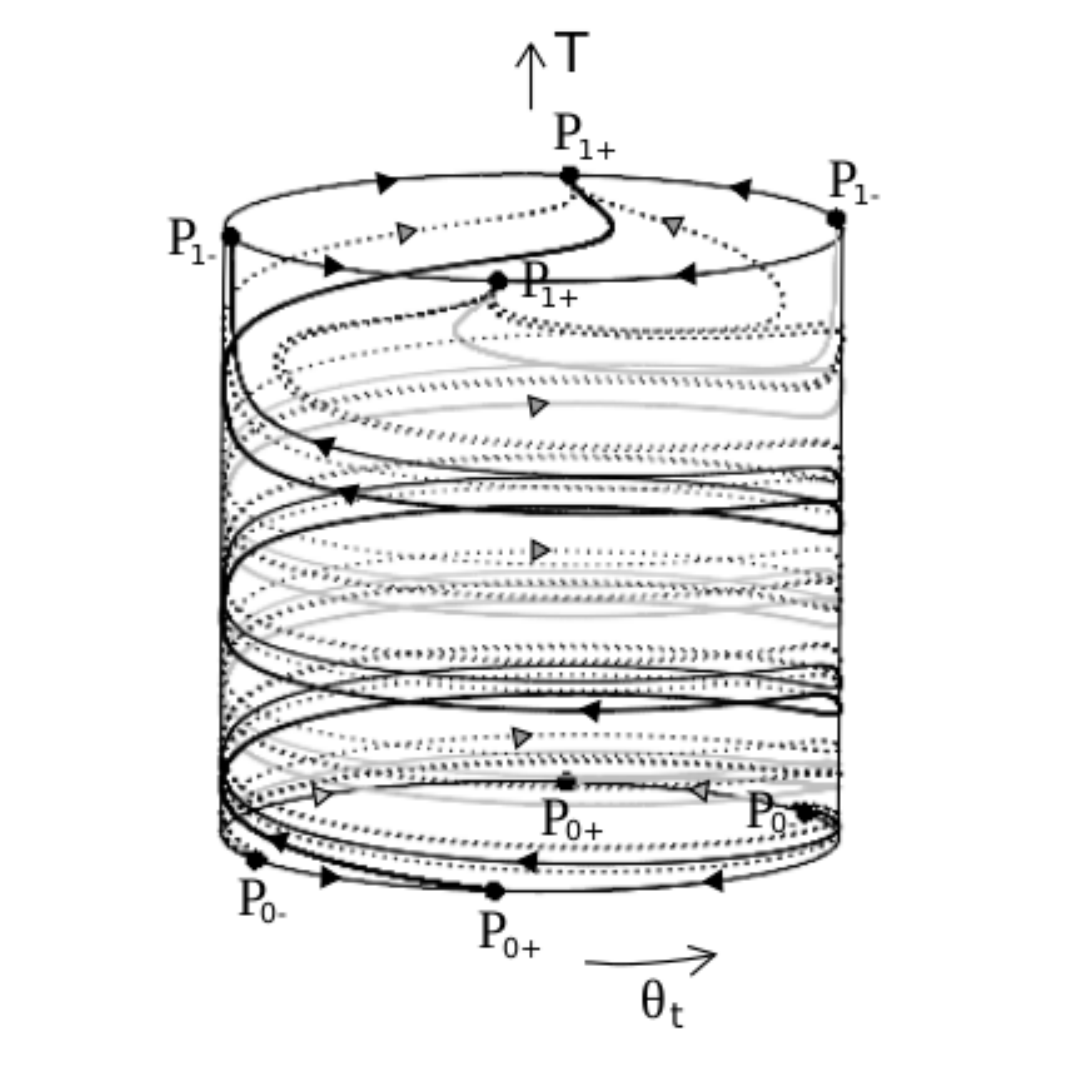}}
		\vspace{-0.5cm}
	\end{center}
	\caption{Solution structure for tensor perturbations for $\Lambda$CDM cosmology
		for a variety of values of $k_0$.}
	\label{fig:Lcdmtensor2}
\end{figure}

The orbit structure for tensor perturbations for $\Lambda$CDM cosmology for
a variety of values of $k_0$ is illustrated in
figure~\ref{fig:Lcdmtensor2}. Note that for large $k_0$, i.e., $k_0\gg 0$, orbits
start to circulate the state space at an intermediate  stage of the evolution.
This regime is approximately
described by the short wavelength limit for dust, but
eventually the cosmological constant starts to dominate
and the future asymptotic limits for the orbits $y_{\mathrm{t}}(N,k_0^2)$
are the same for all values of $k_0$,
since the fixed points at $T=1$ are independent of $k_0$.
%

%

\begin{figure}[ht!]
	\begin{center}
		\subfigure[$k_0=0$]{\label{}
			\includegraphics[width=0.50\textwidth]{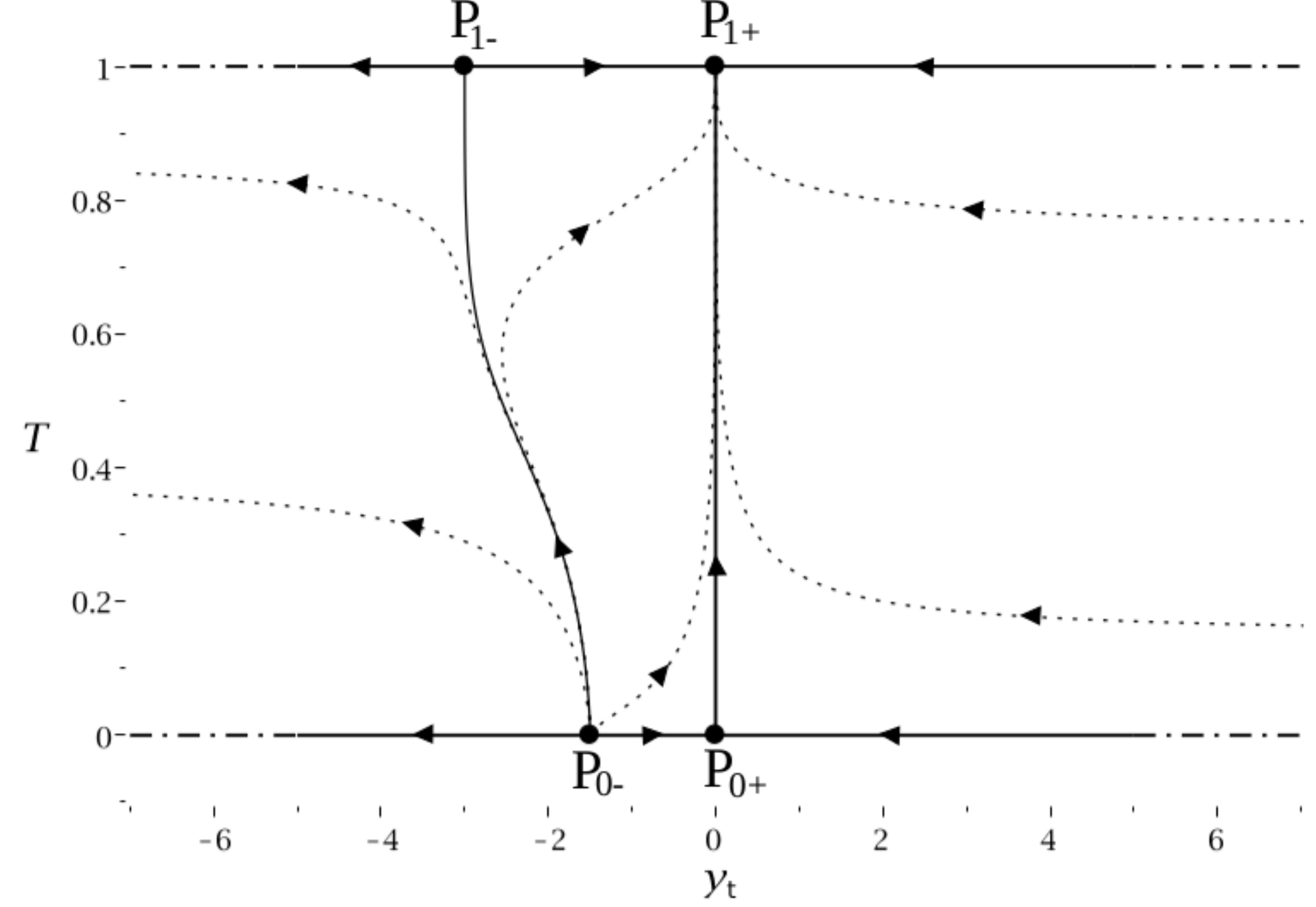}}
		\subfigure[$k_0=0$]{\label{}
			\includegraphics[width=0.45\textwidth]{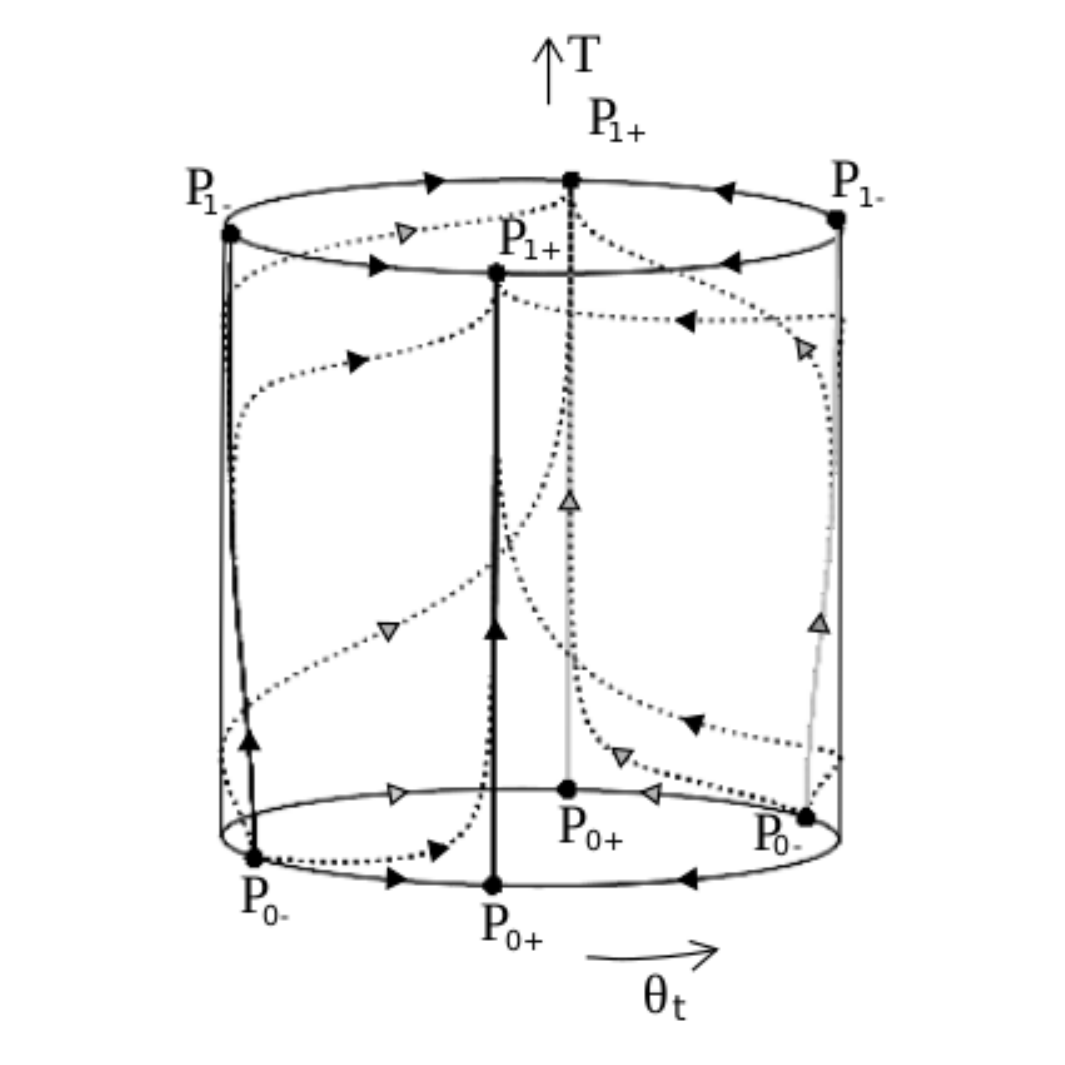}}
		\vspace{-0.5cm}
	\end{center}
	\caption{Solution structure for tensor perturbations for $\Lambda$CDM cosmology
		for the long wavelength limit $k_0=0$.}
	\label{fig:Lcdmtensor}
\end{figure}

The long wavelength limit corresponds to $k_0=0$ and yields an
explicit solution for $y_{\mathrm{t}}$, which is most conveniently
expressed by using
\begin{equation}
\Omega_\Lambda = \frac{\lambda_m x^3}{1 + \lambda_m x^3} = \frac{T^3}{T^3 + (1-T)^3},
\end{equation}
which results in
\begin{equation}
y_{\mathrm{t}} = -\frac{(1-\Omega_{\Lambda})C_+}{C_+ + C_-\Omega_{\Lambda}^{\frac{1}{2}}}.
\end{equation}
In this case there are two special orbits, one with $C_+=0$
going from $\mathrm{P}_{0+}$ to $\mathrm{P}_{1+}$
and one with $C_-=-C_+$, for which $y_{\mathrm{t}} =
 -(1 + \Omega_{\Lambda}^{\frac{1}{2}})$,
going from $\mathrm{P}_{0-}$ to $\mathrm{P}_{1-}$. The structure of the
orbits in the long wavelength limit
is illustrated in figure~\ref{fig:Lcdmtensor}.
%

\section{Measures of anisotropy: the shear and Weyl tensors}\label{sec:shearweyl}

In this paper we have given a global analysis of the evolution of linear
scalar and tensor perturbations of a $\Lambda$CDM universe by formulating
the perturbed Einstein equations as dynamical systems.
In order to place this analysis in perspective we now consider
the family of solutions of the Einstein equations whose matter
content is 
dust and a cosmological constant. These solutions
model a class of universes that generalize the $\Lambda$CDM universe
which is the unique model in this class that describes a completely isotropic
universe with flat spatial geometry. We will thus refer to these universes as
generalized $\Lambda$CDM universes.
Complete isotropy is characterized by the requirement that
the shear tensor $\sigma ^a\!_b$ of the fluid congruence and the
Weyl curvature tensor $C^{ab}\!_{cd}$ are zero, see e.g.~\cite{waiell97}
and references therein. From an observational point
of view one is interested in universes in which the anisotropy is small,
by which one means that the  shear tensor and the Weyl tensor are small
relative to the overall expansion of the universe, described  by the Hubble scalar
$H$. We thus form dimensionless scalars by normalizing the
contracted shear and Weyl tensors\footnote{One can decompose the Weyl tensor
into an electric and magnetic
part relative to a given timelike congruence, with spatial components denoted
$C^{\eta i}\!_{\eta j}$ and $C^{\eta i}\!_{lm}\varepsilon^{lm}\!_j$,
where $\varepsilon_{ijk}$ is the three dimensional
alternating symbol. One can form space-time scalars using $C^{ab}\!_{cd}$
as below,  or spatial scalars using the electric and magnetic parts.}
with an appropriate power of the Hubble scalar $H$:
\begin{equation}
\Sigma^2 = (\sigma^a\!_b \sigma^b\!_a)/H^2,\qquad C^2 = (C^{ab}\!_{cd}C^{cd}\!_{ab})/H^4 .
\end{equation}

There exists a number of results about the evolution of generalized $\Lambda$CDM universes.
First, as regards late times it has been shown that universes in this family that expand indefinitely
approach the de Sitter model for an open set of initial conditions,\footnote{This class of
cosmologies is labelled by eight arbitrary spatial functions, Lim \emph{et al.} (2004)~\cite{limetal04}, page 8.}
in the sense that
\begin{equation} \label{asymp.deSitter}
 \lim_{x\rightarrow \infty}\Sigma^2 =0,\qquad  \lim_{x\rightarrow \infty}C^2 =0,
\qquad \lim_{x\rightarrow \infty}\Omega_m=0.
\end{equation}
Second it has been shown that there is a subset of models which approximate the flat RW (Friedmann-Lema\^{i}tre)
model on approach to the singularity,\footnote{This class of cosmologies is labelled by
three arbitrary spatial functions~\cite{limetal04}, page 11.}
\begin{equation}
 \lim_{x\rightarrow 0}\Sigma^2 =0,\qquad  \lim_{x\rightarrow 0}C^2 =0,
\qquad \lim_{x\rightarrow 0}\Omega_m=1.
\end{equation}
The singularity in these models is referred to as an isotropic singularity
(see Goode and Wainwright (1985)~\cite{goowai85}).
On the other hand a typical model undergoes a more complicated evolution described by BKL
oscillations and possible so-called spike oscillations on approach to the singularity
(see e.g. Uggla (2013)~\cite{ugg13}). Nevertheless, the Hubble-normalized anisotropy scalars $\Sigma^2$
and $C^2$ remain bounded during this process, a result that we will use later in this
section.

In this paper we have illustrated the global solution space of the linearly perturbed
$\Lambda$CDM models using dynamical systems that describe scalar and tensor perturbations
separately. We will now use the shear and Weyl tensors to compare the asymptotic
behaviour of the perturbations at early and late times with the full state space picture
of Einstein's field equations in the Hubble-normalized state space description given
in~\cite{limetal04}, and briefly described above. The purpose with this is to shed light 
on the important issue of assessing the validity of cosmological linear pertubations as 
approximations to solutions of the Einstein field equations.

We refer to~\cite{uggwai11a} for
expressions for the perturbed shear and Weyl tensors.\footnote{See equations
(B35a) and (B41c), which we specialize as follows. We assume
zero anisotropic stress, which implies $\Psi=\Phi$, a flat background ($K=0$)
and for simplicity we exclude the vector mode (${\bf B}_i=0$). Note
that the present $h_{ij} = \sfrac12{\bf C}_{ij}$.}
For the linear scalar and tensor perturbations the perturbed shear tensor is given by
\begin{subequations} \label{shear.Weyl}
\begin{equation}
 \frac{{}^{(1)}\!\sigma^j\!_i}{H}={\cal H}^{-2} {\bf D}^i\!_j({\cal H} V_{\mathrm p} )
 +\sfrac12\partial_N h^i\!_j,
\end{equation}
while the perturbed electric Weyl tensor is given by\footnote{One can give a
similar expression for the perturbed magnetic Weyl tensor, but
we have found that it has similar asymptotic properties as the electric Weyl tensor.}
\begin{equation}
 \frac{{}^{(1)}\!C^{\eta j}\!_{\eta i}}{H^2}=-{\cal H}^{-2} ({\bf D}^i\!_j \psi_{\mathrm p}) +
(\partial_N-{\cal H}^{-2}{\bf D}^2) h^i\!_j .
\end{equation}
\end{subequations}
%

\subsection{Scalar perturbations}

Using the relations~\eqref{gaugtrans}
\begin{equation}
{\cal H}V_{\mathrm p} = \psi_{\mathrm p} +{\cal H}V_{\mathrm c}, \qquad
\psi_{\mathrm p}=-{\cal H}B_{\mathrm c}, \qquad
{\cal H}V_{\mathrm c}=-(1+q)^{-1}\phi_{\mathrm c},
\end{equation}
and the scalar mode extraction operator ${\cal S}^{ij}$ we form the Hubble-normalized shear and electric
Weyl scalars for the scalar perturbations:\footnote{${\cal S}^{ij} = \sfrac32({\bf D}^{-2})^2{\bf D}^{ij}$,
where ${\bf D}^{-2}$ is the inverse spatial Laplacian,
${\bf D}_{ij} := {\bf D}_{(i}{\bf D}_{j)} - \sfrac13 \delta_{ij}{\bf D}^2$, and where spatial
indices are raised with $\delta^{ij}$, see Appendix A in~\cite{uggwai13a}.}
\begin{subequations}
\begin{align}
\Sigma_\mathrm{s} &\equiv \frac{{\cal S}^i\!_j {}^{(1)}\!\sigma^j\!_i}{H}=
-{\cal H}^{-2}\left((1+q)^{-1}\phi_{\mathrm c}+{\cal H}B_{\mathrm c}\right),\\
W_\mathrm{s} &\equiv \frac{{\cal S}^i\!_j {}^{(1)}\!C^{\eta j}\!_{\eta i}}{H^2}=
{\cal H}^{-2}({\cal H}B_{\mathrm c}).
\end{align}
\end{subequations}
In the dynamical systems formulation for the scalar perturbations we have chosen a variable $T$
according to equation~\eqref{choose.T} which yields $1+q = \frac32(1-T)$ and
\begin{equation} \label{H.LCDM}
{\cal H}^{-2} = ({\cal H}_0)^{-2}\left(\frac{1+\lambda_m}{\lambda_m^{1/3}}\right)
T^{1/3}(1-T)^{2/3}.
\end{equation}
One can express $\Sigma_s$ and $W_s$ in the non-reduced state space $({\cal H}B_\mathrm{c}, y_c, T)$,
which yields
\begin{equation}
\Sigma_\mathrm{s} = -{\cal H}^{-2}(1+q)^{-1}({\cal H}B_{\mathrm c})\left(1+q - y_{\mathrm c}\right),\qquad
W_\mathrm{s} = {\cal H}^{-2}({\cal H}B_{\mathrm c}).
\end{equation}
Using the previously obtained asymptotic expressions for $y_\mathrm{c}$, and inserting them into
equation~\eqref{HBc} for ${\cal H}B_\mathrm{c}$ to obtain asymptotic expressions for ${\cal H}B_\mathrm{c}$,
subsequently results in asymptotic expressions when $T\rightarrow 0$ and $T \rightarrow 1$ for $\Sigma_\mathrm{s}$
and $W_\mathrm{s}$. However, since we in the present case have explicit solutions for $\phi_\mathrm{c}$
and ${\cal H}B_\mathrm{c}$, given in equation~\eqref{HBphisol}, we can give explicit expressions for $\Sigma_\mathrm{s}$
and $W_{\mathrm s}$ as functions of $T$:
\begin{subequations} \label{shear.weyl.orbits}
\begin{align}
{\Sigma}_\mathrm{s} &\propto T^\frac13(1-T)^\frac23 C_+ -
\frac32T^{-\frac12}(1-T)(C_+I + C_-),\\
W_{\mathrm s} &\propto \frac32T^{-\frac12}(1-T)(C_+I + C_-),
\end{align}
\end{subequations}
where we recall that $T = \lambda_m x^3/(1 + \lambda_m x^3)$.
It follows that
\begin{equation} \label{sigma.lim.0}
\lim_{x\rightarrow 0}|{\Sigma}_\mathrm{s}|= \infty, \quad\text{if}\quad C_{-}\neq 0,\qquad
\lim_{x\rightarrow 0}{\Sigma}_\mathrm{s}=0, \quad\text{if}\quad C_{-}= 0,
\end{equation}
where the second result follows from $I \propto T^{\frac56}$ when $T \rightarrow 0$.
The Weyl scalar has the same limits. One can also infer the asymptotic rate of
growth/decay of ${\Sigma}_\mathrm{s}$ and $W_{\mathrm s}$ as $x\rightarrow 0$
from equation~\eqref{shear.weyl.orbits}:
\begin{equation} \label{SW.asym.to0, scalar}
{\Sigma}_\mathrm{s},W_{\mathrm s} \propto x^{-3/2},
\quad\text{if}\quad C_{-}\neq 0,\qquad
{\Sigma}_\mathrm{s},W_{\mathrm s}\propto x, \quad\text{if}\quad C_{-}= 0.
\end{equation}
The limits~\eqref{sigma.lim.0}  suggest that the growing mode (the orbit
past asymptotic to the fixed point $P_{0+}$, given by $C_{-}= 0$),
approximates an exact solution with an isotropic singularity. The asymptotic
decay rate in~\eqref{SW.asym.to0, scalar} agrees with the asymptotic results
of Lim \emph{et al.} (2004)~\cite{limetal04} (see equations (4.19) and (5.9)).
On the other hand, the unboundedness of the Hubble-normalized shear and Weyl perturbation
when $C_{-}\neq 0$ indicates that generic  perturbations  become physically unviable,
\emph{i.e.} they do not approximate exact solutions of the Einstein field equations,
when $T$ is sufficiently close to zero
(recall that investigations of the Einstein field equations indicate that
the Hubble-normalized shear and Weyl tensors are expected to be bounded generically).

One can also use~\eqref{shear.weyl.orbits} to determine the
asymptotic behaviour of ${\Sigma}_\mathrm{s}$ and $W_{\mathrm s}$  at late times
($T\rightarrow 1, x \rightarrow \infty$). Since $\lim_{x\rightarrow \infty} I$
is finite it follows that
\begin{equation} \label{SW.infty.scalar1}
\lim_{x\rightarrow \infty} (\Sigma_\mathrm{s},\,W_{\mathrm s})=0,
\end{equation}
for all $C_+,C_-$. Equation~\eqref{shear.weyl.orbits} also gives
the rates of decay along generic orbits asymptotic to the fixed point
$P_{1+} (C_+I_1+C_-\neq0)$:
\begin{equation} \label{SW.infty.scalar2}
\Sigma_\mathrm{s}={\cal O}(x^{-2}), \qquad W_{\mathrm s}={\cal O}(x^{-3}).
\end{equation}
The result~\eqref{SW.infty.scalar1} for perturbations of $\Lambda$CDM  is compatible
with the future asymptotic behaviour of generalized $\Lambda$CDM universes toward a future
de Sitter state, as described by equation~\eqref{asymp.deSitter}, where
equation~\eqref{SW.infty.scalar2} agrees with the asymptotic results of
Lim \emph{et al} (2004)~\cite{limetal04} (see equations (3.24) and (5.8)).

\subsection{Tensor perturbations}

To analyze the tensor perturbations we make the transition~\eqref{h-_ij.Fourier}
to Fourier space and use one of the four real-valued functions $h$ to represent
the perturbation. Thus according to~\eqref{shear.Weyl} the shear will be represented
by $\sfrac12 h'$ and the electric Weyl tensor will be represented by
$\sfrac12(h'+k^2{\cal H}^{-2} h)$, which motivates defining the shear scalar and
electric Weyl scalar for tensor perturbations according to
\begin{equation} \label{sw.tensor}
\Sigma_{\mathrm t}=\sfrac12 h', \qquad
W_{\mathrm t}=\sfrac12(h'+k^2{\cal H}^{-2} h),
\end{equation}
where
\begin{equation}
{\cal H}^{-2} = ({\cal H}_0)^{-2}\left(\frac{1+\lambda_m}{\lambda_m^{1/3}}\right)
\frac{\lambda_m^{1/3}x}{1+\lambda_mx^3},\qquad
T = \frac{\lambda_m^{1/3} x}{1 + \lambda_m^{1/3} x}.
\end{equation}
In order to analyze the behaviour of these scalars along the orbits we first
rewrite the defining equation $y_{\mathrm t}=h'/h$ in the form
$d(\ln h)/dT=y_{\mathrm t}dN/dT = y_{\mathrm t}/(T(1-T))$ and then integrate
to obtain
\begin{equation}
h(T)= h_0\exp\left(\int^T_{{T_0}} \frac{y_{\mathrm{t}}(\tilde{T})}
{\tilde{T}(1-\tilde{T})}d\tilde{T}\right).
\end{equation}
It then follows from~\eqref{sw.tensor} that  $\Sigma_{\mathrm t}$ and $W_{\mathrm t}$
are given as functions of $T$ along the orbits by
\begin{equation} \label{shear.weyl.T}
\Sigma_{\mathrm t}(T)=\sfrac12 h(T)y_{\mathrm t}(T),
\qquad W_{\mathrm t}(T)= \sfrac12 h(T)\left(y_{\mathrm t}(T)+k^2{\cal H}^{-2}(T)\right).
\end{equation}
We can now use the asymptotic expansions for the four
fixed points in the previous section to determine
the asymptotic form of $h(t)$ and hence of $\Sigma_\mathrm{t}$.
We first consider the orbits that are past asymptotic to
$P_{0+}$ and $P_{0-}$  as $T\rightarrow 0 \,( x\rightarrow 0)$,
referring to equations~\eqref{P0+.tensor} and~\eqref{P0-.tensor}:
\begin{subequations}
\begin{alignat}{10}
P_{0+}\!: &\quad y_{\mathrm t} &= {\cal O}(T), &\quad T &\rightarrow 0  & \quad \Rightarrow &\quad
\lim_{T\rightarrow 0}h(T) &\neq 0 &\quad
\Rightarrow &\quad \lim_{x\rightarrow 0}\Sigma_{\mathrm t} &= 0, \\
P_{0-}\!: &\quad y_{\mathrm t} &\approx -\sfrac32, &\quad T &\rightarrow 0 & \quad \Rightarrow &\quad
\lim_{T\rightarrow 0}h(T) &= \infty &\quad
\Rightarrow  &\quad \lim_{x\rightarrow 0}\Sigma_{\mathrm t} &= \infty .
\end{alignat}
\end{subequations}
One can further obtain the leading temporal $x$-dependence, as follows:
\begin{equation}
P_{0+}:\Sigma_{\mathrm t}={\cal O}(x), \quad
P_{0-}: \Sigma_{\mathrm t}={\cal O}(x^{-3/2}),\quad \text{as}\quad x\rightarrow 0.
\end{equation}
Since ${\cal H}^{-2}(0)=0$ it follows from~\eqref{shear.weyl.T} that
$W_{\mathrm t}$ has the same asymptotic behaviour as $\Sigma_\mathrm{t}$.

The above results describe the asymptotic behaviour of the perturbed shear
and electric Weyl scalars near the initial singularity.
The results are the same as for scalar perturbations.
For example, only the single orbit $\mathrm{P}_{0+}\rightarrow \mathrm{P}_{1+} $
in figure~\ref{fig:Lcdmtensor2} and figure~\ref{fig:Lcdmtensor} is compatible with an isotropic
singularity and this orbit is thus analogous to the growing mode orbit
$\mathrm{P}_{0+}\rightarrow \mathrm{P}_{1+}$ for scalar perturbations in
figure~\ref{FigLCDMucg}. Also, analogously with the scalar perturbations, generically
tensor perturbations result in unbounded Hubble-normalized shear and Weyl tensors,
suggesting that the perturbations asymptotically are no longer approximations to the
Einstein field equations when approaching $\mathrm{P}_{0-}$ toward the past.

We now consider the orbits that are future asymptotic to the fixed points
$P_{1+}$ and $P_{1-}$  as $T\rightarrow 1 \,( x\rightarrow \infty)$,
referring to equations~\eqref{P1+.tensor} and~\eqref{P1-.tensor}:
\begin{subequations}
\begin{alignat}{7}
P_{1+}\!: &\quad y_{\mathrm t}\, &=& \, {\cal O}((1-T)^2), &\quad T &\rightarrow 1  &\quad \Rightarrow &\quad
\lim_{T\rightarrow 1}h(T) &\neq 0 &\quad
\Rightarrow &\quad \lim_{x\rightarrow \infty}\Sigma_{\mathrm t} &=0, \\
P_{1-}\!: &\quad y_{\mathrm t}\, &\approx &\, -3, &\quad T &\rightarrow 1 &\quad \Rightarrow &\quad
\lim_{T\rightarrow 1}h(T) &= 0 &\quad
\Rightarrow &\quad \lim_{x\rightarrow \infty}\Sigma_{\mathrm t} &= 0.
\end{alignat}
\end{subequations}
One can further obtain the leading $x$-dependence, as follows:
\begin{equation}
P_{1+}\!: \quad \Sigma_{\mathrm t}={\cal O}(x^{-2}), \qquad
P_{1-}\!: \quad \Sigma_{\mathrm t}={\cal O}(x^{-3}), \quad \text{as}\quad x\rightarrow \infty .
\end{equation}
Again, $W_{\mathrm t}$ has the same asymptotic behaviour as $\Sigma_\mathrm{t}$.
Here there is one difference between the scalar and tensor perturbations: both
$\Sigma_{\mathrm s}$ and $\Sigma_{\mathrm t}$ are ${\cal O}(x^{-2})$
as $x\rightarrow \infty$, while
$W_{\mathrm s}$ is  ${\cal O}(x^{-2})$ and $W_{\mathrm t}$ is ${\cal O}(x^{-3})$.
Since the asymptotic perturbations agree with the asymptotic results
for the full Einstein equations in~\eqref{asymp.deSitter}
one expects that the perturbations will approximate
exact solutions at late times toward the future asymptotic de Sitter state.\footnote{As
regards fluids with rotation, vector perturbations $B_i$ are given by
$B_i= b_i x^{-2}$, where $b_i$ depends on the spatial coordinates only, as follows from
equation~(58a) in~\cite{uggwai11a}. Equations~(58b), (42a), (66d)
in the same reference yields $\tilde{v}_i$, which together with $B_i$ when inserted into
equation~(B.41c) for the vector mode of ${}^{(1)}\!\sigma^i\!_j/{}^{(0)}\!H$ results in that this quantity
is $\propto x^{-\frac32}$ when $x\rightarrow 0$. Hence the vector mode has to be
set to zero in the case of an isotropic singularity. Equation (B.41d) in~\cite{uggwai11a}
for the fluid rotation also results in an initial blow up, unless the vector mode
is set to zero.}

%
%

\section{Concluding remarks}\label{sec:concl}

The purpose of this paper has been to develop a new approach
to using dynamical systems methods to analyze linear perturbations
on a spatially flat RW background.  We decided to use the $\Lambda$CDM model
to illustrate the method  because of its importance in
cosmology and because of its relative mathematical simplicity.
In our approach the state space ${\cal S}$ of the dynamical
system has a product structure
\begin{equation}
{\cal S}={\cal B}\times {\cal P},
\end{equation}
where
${\cal B}$ is the \emph{background state space} which describes
the dynamics of the flat RW background,
and ${\cal P}$ is the \emph{perturbation state space},
which contains the gauge invariant perturbation variables.

The Einstein equations in the RW background give a system
of autonomous differential equations on ${\cal B}$ and the
linearly perturbed Einstein equations give a system
of autonomous differential equations for
the perturbation variables in ${\cal P}$, which involve
the background variables in ${\cal B}$. In this way the dynamics
in the background determine the dynamics of the perturbations. The
advantage of the product structure ${\cal S}={\cal B}\times {\cal P}$ is that
when an orbit on ${\cal B}\times {\cal P}$ is projected onto the background
${\cal B}$ it coincides with an orbit on ${\cal B}$.

A key step is to choose bounded variables
so that the state space is \emph{compactified} and the system of
autonomous differential equations is \emph{regular}. It is
also desirable to take advantage of the fact that the Einstein equations
make it possible to decouple some of the variables, leaving
a \emph{reduced state space} to describe the essential dynamics.

The mathematical simplicity of the $\Lambda$CDM model is reflected in the fact
that it is possible to use only one background variable
and one perturbation variable to describe the essential dynamics.
In other words both ${\cal B}$ and ${\cal P}$ are one dimensional
spaces. For example, in the case of scalar perturbations in the uniform
curvature gauge we represented ${\cal B}$ as the unit line segment $L$
with $T=\Omega_{\Lambda}$ as the background variable,
and ${\cal P}$ as the circle ${S}^1$ described by the angular
variable $\theta_{\mathrm c}$ . The state space is
thus ${\cal S}={L}\times {S}^1,$ which is a finite segment of a cylinder.
Because the state space is bounded we were able to give a global description of
the dynamics, in particular the behaviour at early and
late times and the evolution at intermediate stages that may be
of physical interest. In addition  the differential equations,
using $e-$fold time $N$ are well-suited for performing numerical simulations.

In future papers we will show how to obtain \emph{reduced and compactified}
product state spaces ${\cal B}\times {\cal P}$ with systems of \emph{regular
differential equations} for scalar field models and models with multiple sources.
As a first step  we will consider the simplest case, namely
a minimally coupled scalar field with exponential potential, for which
${\cal B}$ is two dimensional  and ${\cal P}$ is one dimensional. This
case will provide the basis for the generalization to the case of
a scalar field with more general potentials, which requires that ${\cal B}$
is three dimensional. This will also illustrate how one can organize perturbation
theory into hierarchical structures where simpler models act as building
blocks for more complicated ones.

\subsection*{Acknowledgments}
AA is funded by the FCT grant SFRH/BPD/85194/2012, and supported by the project (GPSEinstein) PTDC/MAT-ANA/1275/2014,
and CAMGSD, Instituto Superior T{\'e}cnico by FCT/Portugal through UID/MAT/04459/2013. Furthermore, AA thanks the
warm hospitality of Karlstad University. CU would like to thank the CAMGSD, Instituto Superior T{\'e}cnico in Lisbon
and the University of Waterloo, Canada, for kind hospitality.

\bibliographystyle{plain}
\bibliography{../Bibtex/cos_pert_papers}

\end{document}